\documentclass[usenatbib]{mn2e}
\usepackage{amsmath}
\usepackage{graphicx}
\usepackage{pgfplots}
\usepackage{cuted}
\pgfplotsset{compat=1.5}

\def\lesssim{\mathrel{\hbox{\rlap{\hbox{\lower4pt\hbox{$\sim$}}}\hbox{$<$}}}}
\def\gtrsim{\mathrel{\hbox{\rlap{\hbox{\lower4pt\hbox{$\sim$}}}\hbox{$>$}}}}

\newcommand{\mamo}[1]{\mbox{$#1$}}
\newcommand{\unit}[1]{\ifmmode \:\mbox{\rm #1}\else \mbox{#1}\fi}
\newcommand{\sbr}[1]{_{\rm #1}}

\newcommand{\expec}[1]{\mamo{\left\langle #1 \right\rangle}}
\newcommand{\std}[1]{\mamo{\sigma\left[#1\right]}}

\newcommand{\var}[1]{\mamo{\sigma^2 \left[#1\right] }}

\newcommand{\g}{g}
\newcommand{\Seclabel}[1]{\label{sec:#1}}

\newcommand{\Applabel}[1]{\label{sec:#1}}
\newcommand{\Eqlabel}[1]{\label{eq:#1}}
\newcommand{\Figlabel}[1]{\label{fig:#1}}
\newcommand{\Tablabel}[1]{\label{tab:#1}}
\newcommand{\Secref}[1]{Section~\ref{sec:#1}}

\newcommand{\Appref}[1]{Appendix~\ref{sec:#1}}
\newcommand{\Eqref}[1]{Equation~(\ref{eq:#1})}
\newcommand{\Figref}[1]{Fig.~\ref{fig:#1}}
\newcommand{\Tabref}[1]{Table~\ref{tab:#1}}

\title{The Effects of Calibration on the Bias of Shear Measurements}
\author[B.R. Gillis \& A.N. Taylor]{Bryan R. Gillis$^{1}$\thanks{E-mail: b.gillis@roe.ac.uk}, A.N. Taylor$^{1}$ \\ \nonumber
$^{1}$School of Physics and Astronomy, University of Edinburgh, Royal Observatory Edinburgh, Edinburgh, EH9 3HJ, United Kingdom.}

\begin{document}

\date{13 Sep 2018}

\pagerange{\pageref{firstpage}--\pageref{lastpage}} \pubyear{2018}

\maketitle

\label{firstpage}

\begin{abstract}

\noindent Forthcoming large-scale surveys will soon attempt to measure cosmic shear to an unprecedented level of accuracy, requiring a similarly high level of accuracy in the shear measurements of galaxies. Factors such as pixelisation, imperfect point-spread function (PSF) correction, and pixel noise can all directly or indirectly lead to biases in shear measurements, and so it can be necessary for shear measurement methods to be calibrated against internal, external, or simulated data to minimize bias. It is thus important to understand the nature of this calibration. In this paper, we show that a typical calibration procedure will on average leave no residual additive bias, but will leave a residual multiplicative bias. Additionally, the errors on the post-calibration bias parameters will be changed, and on average increased, from the errors on the pre-calibration measurements of these parameters, but that this is generally worth the benefit in decreasing the expected value of the multiplicative bias. We find that in most typical cases, it is worthwhile to apply a first-order bias correction, while a higher-order bias correction is only worthwhile for methods with intrinsically high multiplicative bias ($>10$ per cent) or when the simulation size is very small ($<10^6$ simulated galaxies).

\end{abstract}


\section{Introduction}

Cosmic shear measurements have the potential to aid in constraining the dark energy equation of state through its sensitivity to the geometry of the universe and the growth of structure throughout its history \citep{AlbBerCah06,TayKitBac07,AmeAppBac13}. While its sensitivity to structure shows a large degeneracy between $\Omega_m$ and $\sigma\sbr{8}$, this degeneracy is orthogonal to that from cosmic microwave background (``CMB'') measurements \citep[eg.][]{HeyGroHea13}, allowing the degeneracies to be broken through the combination of the two methodologies. This helps place tighter constraints on the allowable values for these parameters, indirectly tightening the constraints on other cosmological parameters as well.

Cosmic shear is measured from the coherent alignment of galaxies due to gravitational lensing distortion and how this relates to their relative positions and redshifts \citep{Gun67,BacRefEll00,KaiWilLup00,VanMelErb00,WitTysKir00}. The distortion in galaxy shapes due to cosmic shear is $\lesssim 1$ per cent, and measurements of it are intrinsically noisy due to the unlensed shapes of galaxies, sky background noise, and Poisson pixel noise. They also face difficulty due to smearing of the image from the telescope's point-spread function, pixelisation, the presence of blended objects, and irregular galaxies in the dataset, among other factors. The wealth of data available means that a large statistical error on the shear measurements of individual galaxies is acceptable, as this can be reduced through the large number of galaxy images.

It is not, however, possible to remove a bias in shear measurements through simply gathering a large amount of data. As weak lensing surveys get larger and the statistical error on shear measurements is minimized, the effects of measurement bias will begin to dominate the measured values, and this bias can carry through to affect the estimates of cosmological parameters \citep{HutTakBer06,AmaR08,KitTayHea08,MasHoeKit13,TayKit16}. It thus becomes necessary to either develop shear measurement methods with minimal bias or to measure and correct for the biases of methods. In either case, it is necessary to test the method against a large set of calibration data, either to demonstrate that the method is unbiased at a certain threshold or to accurately determine the bias so that it can be corrected. In either case, we must consider what the bias distribution will be after a correction or with no correction.

The typical amount of bias varies greatly depending on the shear-estimation method used and the circumstances of its use (e.g. ground-based versus space-based observations, high-signal-to-noise galaxies versus low signal-to-noise galaxies, etc.). For instance, the third Gravitational Lensing Accuracy Testing challenge \citep[``GREAT3''][]{ManRowArm15} found, through testing against simulations with known shears, that one implementation of the Kaiser-Squires-Broadhurst \citep[``KSB''][]{KaiSquBro95} method (which is still commonly used due its speed) has additive bias of order $6\times10^{-4}$ and multiplicative bias of order $3 \times 10^{-2}$ for ground-based data, but this bias increases by an order of magnitude for space-based data. More modern methods such as Bayesian Fourier Domain \citep[``BFD''][]{, BerArm14,BerArmKra16} and Metacalibration \citep{SheHuf17} claim lower or even zero bias intrinsic to the methods, but can still manifest multiplicative bias at the per-cent level due to external factors such as object selection and crowding \citep[see e.g.][]{JarSheZun16,FenHerHoe17}. We thus decide in this paper to investigate the calibration of multiplicative biases ranging from $1$ to $20$ per cent.

Since it is only recently and in the near future that lensing data of sufficient volume will be available to require precise calibration, there is not yet consensus on the best way to do this. Favouring the calibration approach, \citet{HoeVioHer17} argues that ``a known bias can be incorporated as part of an empirical calibration step, thus reducing the `effective' residual bias,'' but does not detail how this calibration step would be performed. In analysing the lensing measurements from the Canada-France-Hawaii Telescope Lensing Survey \citep[``CFHTLenS''][]{ErbHilMil13}, \citet{MilHeyKit13} used a forward-modelling approach where the measured bias is used in a correction term for the two-point shear correlation function. For the Kilo-Degree Survey \citep[``KiDS''][]{FenHerHoe17,HilVioHey17}, the bias is calculated for bins of size and luminosity, and then corrected at the catalogue level for each galaxy in each bin. It is this final approach that we analyse in this paper, looking at various methods in which the residual bias can be reduced.

As shear-measurement algorithms can be slow, sometimes requiring seconds per galaxy on average for model-fitting methods, practical limitations will exist on the possible size of a calibration dataset. This will limit the accuracy of the bias measurements, and thus any correction based on these measurements will likely have some residual bias. It is important to properly calculate the expected value and scatter of this residual bias, so that this source of uncertainty can be properly propagated through to the uncertainties on the measured cosmological parameters. This propagation was investigated in \citet{TayKit16}, where the authors assumed that there would be no residual bias on average after calibration. In this paper, we test that assumption, and we calculate the expected scatter in the residual bias after calibration and look into the trade-off between minimizing the expected value of the residual bias and the increase in its scatter due to calibration. We also look into how we can determine an unbiased estimate of the mean residual bias using only measured values, and we show calculations for how large a calibration dataset must be to achieve a desired maximum bias.

\citet{FenHerHoe17} calibrated using a simulation with $\sim 10^7$ galaxies, applying shape-noise cancellation to achieve the equivalent statistical power of using $\sim 10^8$ galaxies, which they split into $\sim 200$ bins of size and signal-to-noise, resulting in the effective statistical power of $\sim 5\times10^5$ galaxies per bin. For this paper, we thus choose to look at the statistics that result from using a simulation of $10^6$ galaxies to approximately this statistical power. We also choose to test with a simulation size of $10^4$ to help demonstrate how certain statistics scale with the simulation size. We present here a simple scenario in which no weighting is applied to galaxies and the bias correction is applied on a per-pixel basis.

In \Secref{calcs} of this paper, we present our calculations of the expected distribution of residual bias after calibration. In \Secref{illustration}, we present an illustration of how bias distributions are changed as a result of calibration and discuss how it depends on the initial bias. In 
\Secref{projections}, we calculate projects for the expected residual bias distributions after calibration, depending on the design and size of the calibration dataset, and we calculate the needed size of the dataset to reduce the typical bias below desired thresholds. Finally, we conclude in \Secref{conclusions}.

\section{Bias Correction}
\Seclabel{calcs}

\subsection{Assumptions}

In order to simplify the calculations in this paper, we make a few assumptions about the nature of the calibration problem, which can be loosened in future work:
\begin{enumerate}
	\item The error on shear measurements can be represented as a Gaussian deviate.
	\item The domains for the bias parameters and the shear variables span all real numbers.
	\item The two components of the shear polar can be treated independently.
\end{enumerate}
The assumption of a uniform prior can be loosened if the priors take the forms of Gaussian probability distributions, as this will only affect the measured bias parameters and their errors in our calculations. In general, our calculations are actually not strongly dependent on the assumption of Gaussianity, with the notable exceptions of the higher-order terms in \Eqref{post_correction_vars}, \Eqref{second_known_bias}, and \Eqref{second_correction_vars}.

While the assumption of infinite domain is not true for standard ellipticity and shear parametrisations, which are constrained to the unit circle, measurements of these values are often allowed infinite domain in order to avoid truncating the error distribution, and so this assumption is not unreasonable.

In this paper, we choose to express shear measurements as the reduced shear $g=\gamma/(1-\kappa)$, where $\gamma$ is the shear and $\kappa$ is the magnification, although the calculations here remain valid for other parametrisations, as long as the bias takes the same form.

We make the assumption that the two components of the shear polar can be treated independently. This is not necessarily going to be the case for all shear-measurement algorithms, but a full treatment of this possibility is beyond the scope of this paper. Throughout the remainder of the paper, we consider only a single component of shear, and assume that our results can be applied to both components independently.

\subsection{Approaches to Calibration}

In this section we discuss approaches to calibration from a frequentist perspective. For a Bayesian analysis, see \Secref{bayesian_methods}.

Let us define $\hat{x}$ as the measured value of any quantity $x$, $\std{x}$ as the uncertainty in the measured value of $x$, and $\delta_x$ as a single random realisation of a Gaussian random variable with zero mean and standard deviation $\std{x}$.

For a measurement $\hat{\g}$ of reduced shear $\g$, we express its bias in terms of additive and multiplicative components $m$ and $c$. It is thus related to the actual value of $\g$ through
\begin{equation}
	\hat{\g} = (1+m)\g+c + \delta_{\g}\mathrm{,}
\end{equation}
where $\delta_{\g}$ is the bias-free random offset of an individual measurement. This can be rearranged as
\begin{equation}
	\g = \frac{\hat{\g}-c-\delta_{\g}}{1+m}\mathrm{.}
	\Eqlabel{bias_rearranged}
\end{equation}
In reality, we do not know the actual values of $m$ and $c$. Rather, we have measured values $\hat{m}$ and $\hat{c}$, which are related to $m$ and $c$ through
\begin{align}
	\hat{m} &= m+\delta_m \;\mathrm{and}\;\hat{c} = c+\delta_c \mathrm{.}
\end{align}
The intuitive first-order corrected estimate for $\g$ would be
\begin{equation}
	\hat{\g}^p = \frac{\hat{\g} - \hat{c}}{1 + \hat{m}}\mathrm{.}
	\Eqlabel{intuitive_bias_correction}
\end{equation}
This is the form of the correction applied by \citet{FenHerHoe17}. However, this estimate has the problem that it is dividing by a noisy value, which results in a non-Gaussian distribution and formally divergent variance, making it difficult to analyse. If the probability that $1+m$ will approach zero is negligible though, the distribution of $1/(1+\hat{m})$ can be approximated through e.g. a Taylor expansion. This shows that it if $\hat{m}$ is unbiased (as is the case resulting from a simple linear regression model), then $1/(1+\hat{m})$ will be biased high proportional to $\var{m}$ at lowest order. This leaves us with the scenario where we know that this correction is biased, but it has statistics which make it difficult to analyse in-depth, and so it is worth looking for alternatives which might be more tractable.

Since the problem with $\hat{\g}^*$ is that it involves dividing by $1/(1+\hat{m})$, this could be avoided if it were possible to instead determine an unbiased estimator for $1/(1+m)$. The intuitive method to do this would be performing a linear regression of \Eqref{bias_rearranged} and using the slope of this regression, but unfortunately this does not work. The reason for this is that a linear regression only gives an unbiased estimate of the slope if there are only errors in the values of the dependent variable. In the case of shear measurement, the input shear $g$ can be known exactly in simulations, but there will inevitably be errors in the measured values $\hat{g}$, which allows us to only obtained an unbiased estimate of the slope of $\hat{g}$ versus $g$, and not vice-versa. This issue is discussed further in \Secref{machine_learning}.

Given that \Eqref{intuitive_bias_correction} requires the use of approximations to analyse its behaviour, an alternative approach is to instead use these approximations in the correction itself. That is, use a correction which includes the first few terms of the Taylor expansion of \Eqref{intuitive_bias_correction}:
\begin{equation}
	\Eqlabel{bias_correction}
	\hat{\g}' = \left(\hat{\g}-\hat{c}\right)\left(1-\hat{m}+\hat{m}^2\right)\mathrm{.}
\end{equation}
This correction will be slightly more biased than \Eqref{intuitive_bias_correction}, but it has the significant benefits that, since it doesn't involve any division, its statistics are significantly more well-behaved: Its variance is formally finite, and its expected value can be calculated without any assumption of Gaussianity. The biggest drawback to this correction is that its bias will have an $m^3$ term which cannot be reduced through a larger sample size, but if this an issue, a similar, less-biased correction can be generated either through using more terms of the series expansion or through iteration.

For this paper, we analyse the statistics of the bias correction in \Eqref{bias_correction}, since it is much more well-behaved than \Eqref{intuitive_bias_correction} and the benefits of being able to better understand the statistics of the correction outweigh the drawback of the slightly increased bias.

\subsection{Calibration Statistics}
\Seclabel{calib_stats}

We wish to determine the expected bias and uncertainty in bias of $\hat{\g}'$. Let us start by replacing $\hat{m}$ and $\hat{c}$ in \Eqref{bias_correction} with $m+\delta_m$ and $c+\delta_c$, giving us
\begin{equation}
	\hat{\g}' = \left(\hat{\g}-(c+\delta_c)\right)\left(1-(m+\delta_m)+(m+\delta_m)^2\right)\mathrm{.}
\end{equation}
Substituting $\hat{\g}=(1+m)g+c+\delta_{\g}$ into this and expanding, we get
\begin{align}
	\Eqlabel{g_bias_first_step}
	\hat{\g}' &= (1 - \delta_m  + m\delta_m + \delta_m^2 + m^3 + 2m^2\delta_m + m\delta_m^2) \g \\ \nonumber
		 &\;\;\;\;{}+(\delta_{\g}-\delta_c)(1 - m - \delta_m + m^2 + 2m\delta_m + \delta_m^2)\mathrm{.}
\end{align}
We make the assumption that $\delta_{\g}$, $\delta_c$, and $\delta_m$ are independent\footnote{This is reasonable if we assume that $m$ and $c$ are estimated using a simple linear regression on simulated data where the mean applied shear is zero, but may not be the case universally.}. We can then use the properties
\begin{align}
	\Eqlabel{expec_values}
	\expec{\delta_{\g}} &= \expec{\delta_c} = \expec{\delta_m} = 0 \mathrm{, and} \\ \nonumber
	\expec{\delta_m^2} &= \var{m}
\end{align}
to find the expectation
\begin{equation}
	\Eqlabel{gp_expec_value}
	\expec{\hat{\g}'} = (1 + \var{m} + m\var{m} + m^3)\g\mathrm{.}
\end{equation}
An expanded calculation of this can be found in \Appref{exp_calcs}.

This means that we expect $\hat{\g}'$ to have an expected multiplicative bias of $(1 + \var{m} + m\var{m} + m^3)$ and no expected additive bias. We can express this result as:
\begin{align}
	\Eqlabel{post_correction_known_biases}
	\expec{m'} &= \var{m}(1 + m) + m^3\mathrm{,} \\ \nonumber
	\expec{c'} &= 0\mathrm{.}
\end{align}
This result shows that the residual bias has both a component which scales with $\var{m}$, which comes from the accuracy to which $m$ is measured, and a component which scales as $m^3$. This latter component is due to the fact that we used only a 2nd-order expansion of $g$ in \Eqref{bias_correction}; if we had used a higher-order expansion, the remaining term here would have a correspondingly higher power, but it would also introduce more terms which scale with $\var{m}$ multiplied by some power of $m$. For instance, if we had instead expanded to the fourth power in $\hat{m}$, the bias would instead be:
\begin{align}
	\Eqlabel{post_correction_alternative_known_bias}
	\expec{m^*} &= \var{m}(1 - 2m - 3m^2) - m^4\mathrm{.} \\ \nonumber
\end{align}
There is thus a trade-off in the terms here. If we wish to reduce the bias further, it will be more productive to apply corrections iteratively, which we discuss later in this section. However the corrections are applied, it will be impossible to eliminate some dependence on $m$, meaning that methods with innately large $m$ will be more difficult to calibrate.

We must also consider the uncertainty in the bias of $\hat{\g}'$. To do this, we start by calculating its variance. If we express $\hat{\g}'$ as
\begin{equation}
	\Eqlabel{no_cov_variance_base}
	\hat{\g}' = (1+m')\g + c' + \delta_{\g}'\mathrm{,}
\end{equation}
then the variance of this is
\begin{align}
	\Eqlabel{no_cov_variance}
	\var{\hat{\g}'} &= \var{m'}\g^2 + \var{c'} + \var{\g'}\mathrm{.}
\end{align}
We can therefore determine $\var{m'}$, $\var{c'}$, and $\var{g'}$ by calculating the variance of our expression for $\hat{\g}'$ given above. As before, using our definitions of $\delta_{\g}$, $\delta_c$, and $\delta_m$ as realisations of Gaussian random distributions, we have
\begin{align}
	\var{\delta_{\g}} &= \var{\g}\mathrm{,} \\ \nonumber
	\var{\delta_c} &= \var{c}\mathrm{,} \\ \nonumber
	\var{\delta_m} &= \var{m}\mathrm{,}
\end{align}
and we can use the property of Gaussian distributions that
\begin{equation}
	\Eqlabel{Gaus_dev_combination}
	\var{a\delta_m + b\delta_m^2} = a^2\sigma^2\left[m\right] + 2b^2\sigma^4\left[m\right]\mathrm{.}
\end{equation}

This gives us the resulting variances on $m'$, $c'$, and $\g'$. To lowest order, these are
\begin{align}
	\Eqlabel{post_correction_vars}
	\var{m'}  &\approx \var{m}(1-2m-3m^2+2\var{m})\mathrm{,} \\ \nonumber
	\var{c'}  &\approx \var{c}(1-2m+3m^2+3\var{m})\mathrm{,} \\ \nonumber
	\var{\g'} &\approx \var{\g}(1-2m+3m^2+3\var{m})\mathrm{.}
\end{align}
The full versions of these can be found in \Appref{exp_calcs}. Note that due to the higher-order terms in $\delta_m$ in the above calculation, the resulting distributions for $m'$ and $c'$ will not be Gaussian. For low $\std{m}$, this will be a small deviation, however, and may be negligible.

If the expected multiplicative bias is dominated by the $\var{m}$ term, then we can see that this correction will have reduced the known bias component to be of comparable magnitude to the unknown component, and further corrections will therefore give negligible benefit, as the standard deviation of the unknown bias components will remain dominant. If the $m^3$ term in the bias is dominant, however, further correction may be necessary. To eliminate all known terms of the bias, we can use the correction
\begin{align}
	\Eqlabel{second_bias_correction}
	\hat{\g}'' &= \hat{\g}'\left( 1 - \var{m} + 2\hat{m}\var{m} - \hat{m}^3 \right)\mathrm{.}
\end{align}
This second-order correction will then have a known bias of
\begin{align}
	\Eqlabel{second_known_bias}
	\expec{m''} &\approx \var{m}\bigg[\var{m} + 3m^2\bigg] - m^6 \mathrm{.}
\end{align}
The full version of this can be found in \Appref{exp_calcs}. Note that this calculation relies more heavily on the assumption of Gaussianity in $\delta_m$, since it requires the assumption that $\expec{\delta_m^4} = 2\sigma^4\left[m\right]$, which is only guaranteed for Gaussian distributions.

The variance on $m''$ will be to lowest order
\begin{align}
	\Eqlabel{second_correction_vars}
	\var{m''}  &\approx \var{m}(1-2m+3m^2+2\var{m})\mathrm{.}
\end{align}
The notable change here is that the $3m^2$ term is now positive, and this variance will be consistently higher than the variance on $m'$. It also has many more higher-order terms not shown here, many of which depend heavily on the assumption of Gaussianity. This means that there is a trade-off in choosing the order of calibration. A higher-order calibration will decrease the known component of the bias, but at the expense of increasing the variance in the unknown component and the sensitivity to non-Gaussianity. The calibration order should thus be determined based on a balance of these factors, taking into account the expected pre-calibration $m$ of the method being calibrated.

\section{Simulated Bias Correction}
\Seclabel{illustration}

The effects of calibration on the bias parameters can be illustrated through performing a calibration of mock data with known biases. We start by generating multiple sets of mock data with the following representative parameters: $\std{\g} = 0.25$, which represents the typical per-galaxy measurement error, including the effects of shape noise; $\sigma_{\g} = 0.03$, which represents the Gaussian scatter in intrinsic shear;	$n = 10^4,\; 10^6$ mock shears in each datasets;	$m = -0.2,\; -0.1,\; -0.01,\; 0,\; 0.01,\; 0.1,\; 0.2$;	$c = -0.1,\; 0,\; 0.1$. We generate separate datasets for all combinations of $n$, $m$, and $c$, all using the same random seeding.

We generate multiple realisations for each dataset, and for each realisation we perform a linear regression to determine the measured $\hat{m}$ and $\hat{c}$ values and their errors. The realisation is then corrected using either our first- or second-order correction with these values. We then perform a second linear regression to determine the known components of the residual bias parameters. We also apply a bias correction using the measured $\hat{m}$ and $\hat{c}$ values to a set of noise-free data (where $\std{\g}$ is identically zero), and perform a linear regression on this corrected dataset as well. This gives us the actual $m$ and $c$ values after correction.

\begin{figure*}
	\includegraphics[scale=0.53]{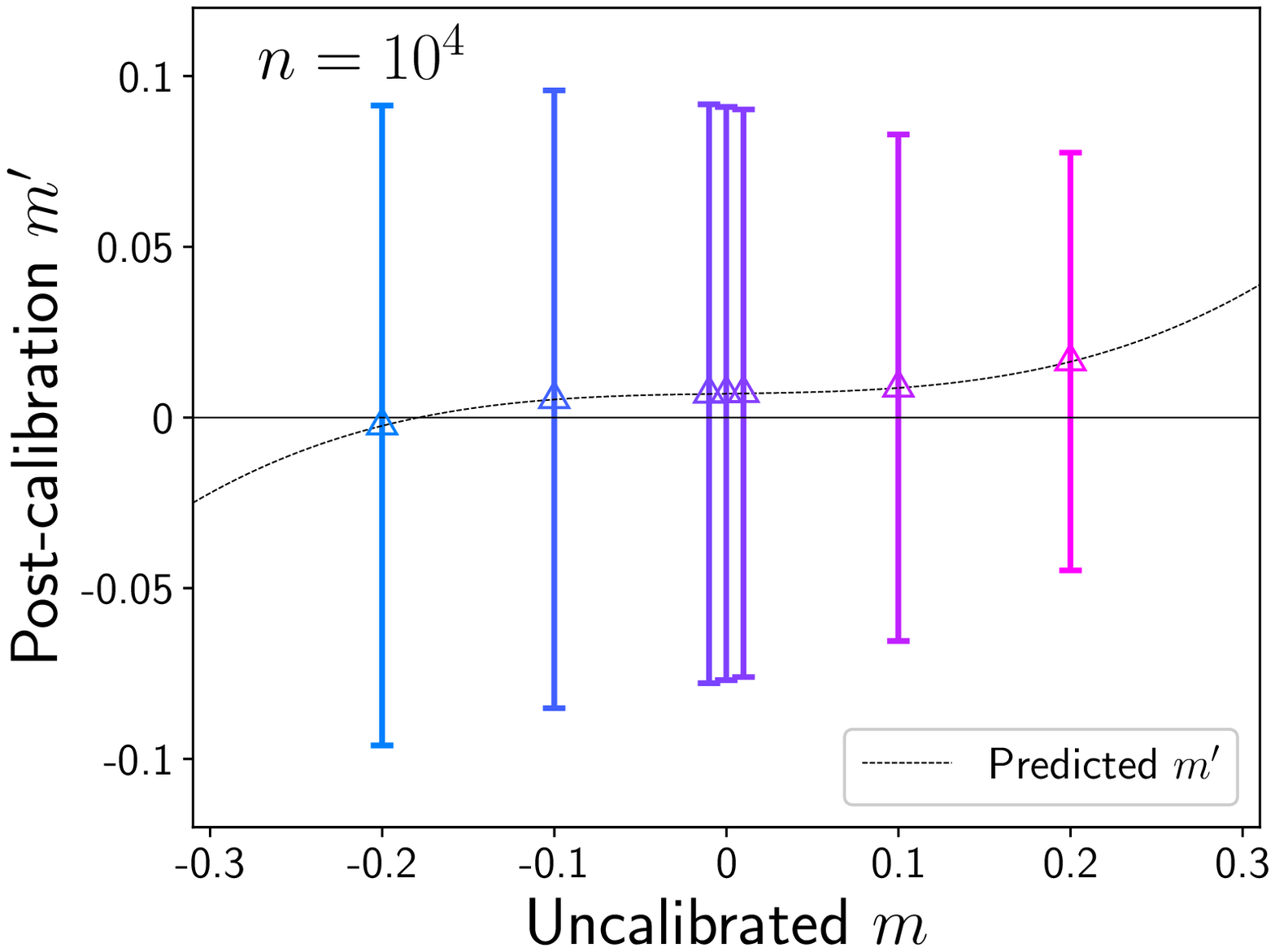}
	\includegraphics[scale=0.53]{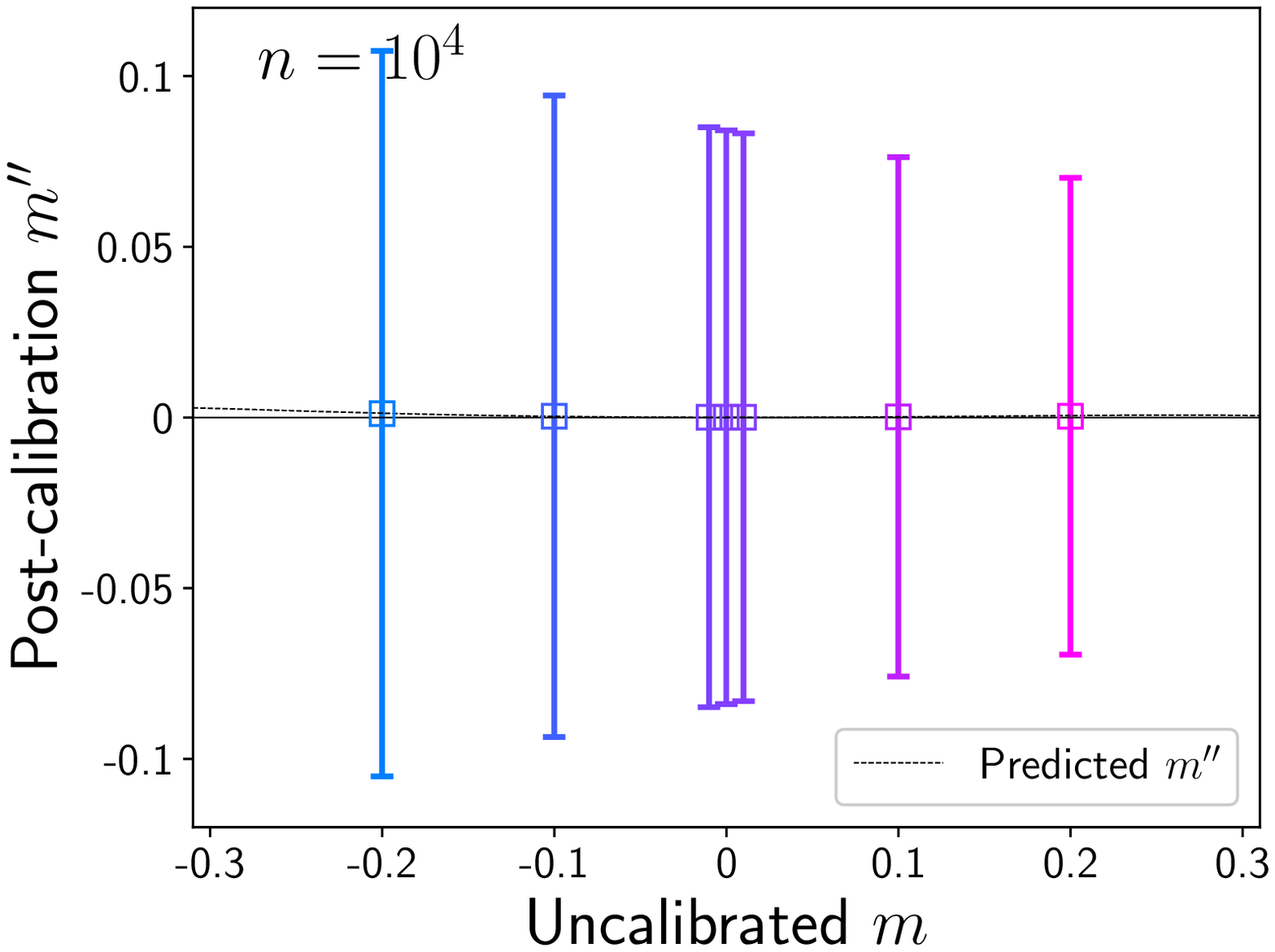}
	\caption{An illustration in the effects of calibration on the $m$ bias parameter, using multiple realisations of datasets of $10^4$ mock shear values. The horizontal axes show the measured pre-calibration values, and the vertical axes show the post-calibration values for a first-order correction ($\g'$, left panel), and a second-order correction ($\g''$, right panel). The markers are placed at the mean value over all test runs, and the errorbars indicate the expected $1\sigma$ scatter of $m$ for a simulation of this size.}
	\Figlabel{bias_correction_illustration_1e4}
\end{figure*}

\begin{figure*}
	\includegraphics[scale=0.53]{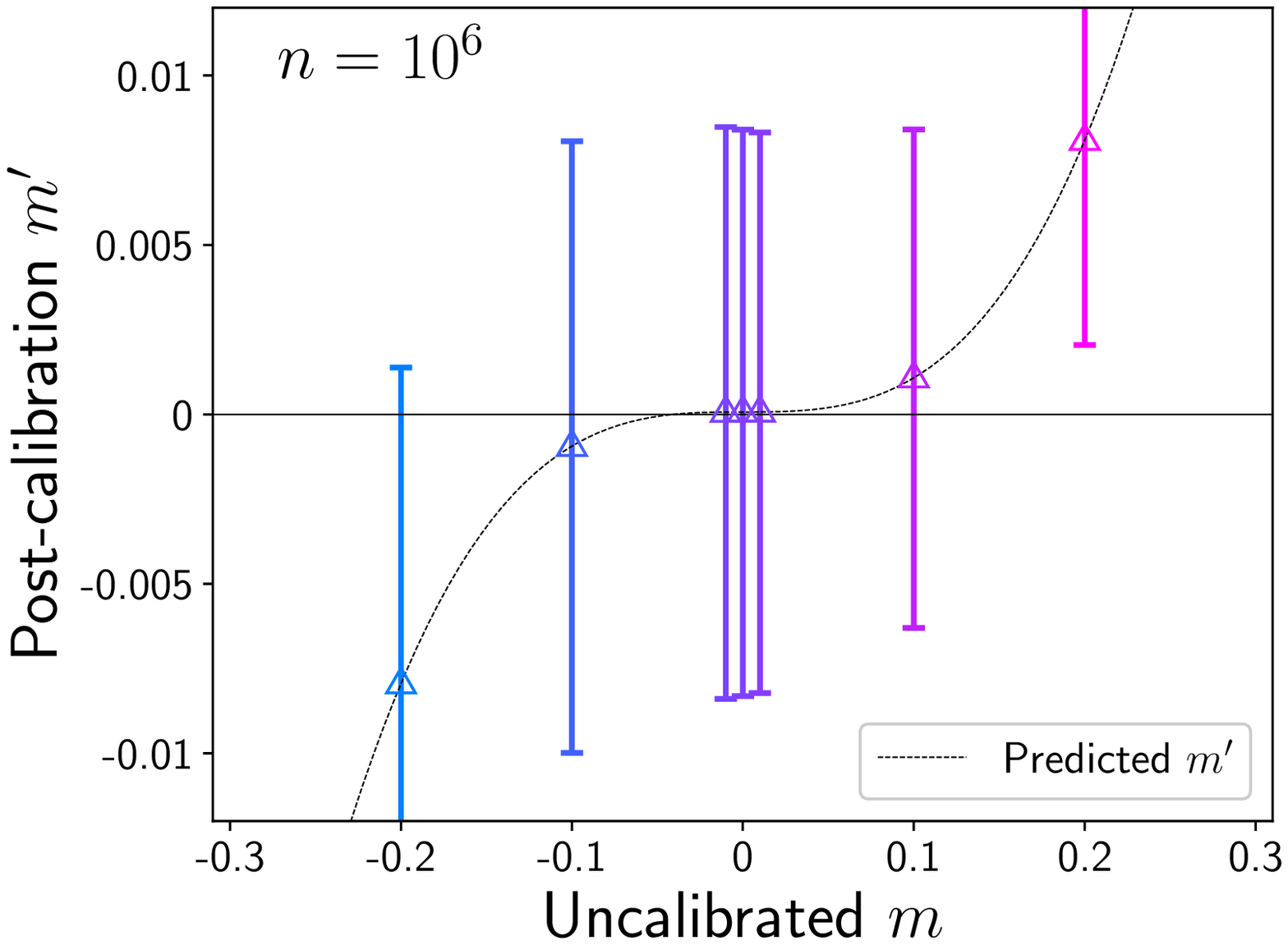}
	\includegraphics[scale=0.53]{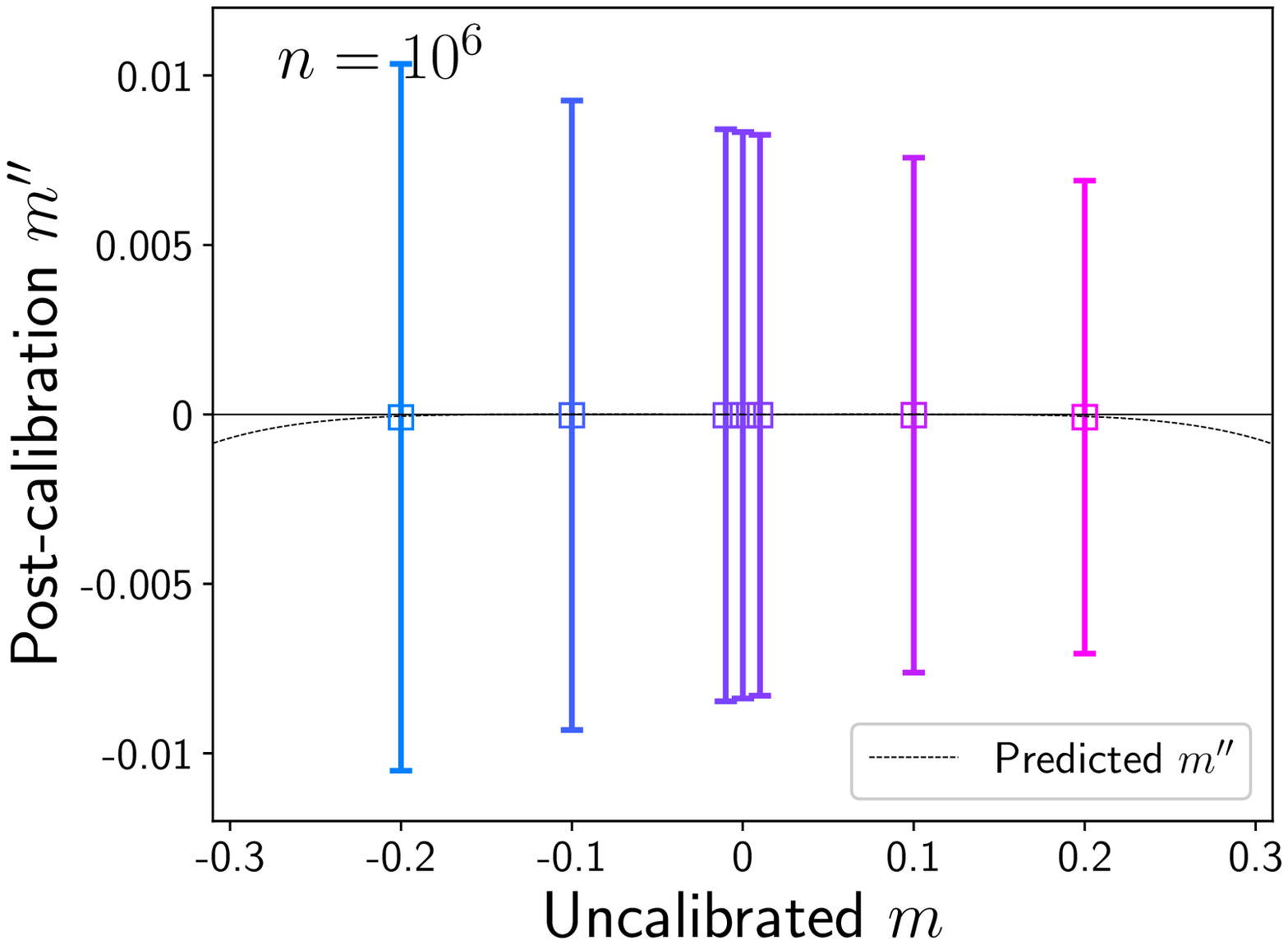}
	\caption{As \Figref{bias_correction_illustration_1e4}, except with $10^6$ mock shear values in each dataset. Note that the vertical axis here, corresponding to post-calibration $m$ values, is at one-tenth the scale of that in \Figref{bias_correction_illustration_1e4}.}
	\Figlabel{bias_correction_illustration_1e6}
\end{figure*}

We illustrate our results in \Figref{bias_correction_illustration_1e4} and \Figref{bias_correction_illustration_1e6}, showing the post-calibration $m$ biases versus the measure pre-calibration values for the first-order correction (left panels) and the second-order correction (right panels), along with the post-calibration values of $m$ predicted by \Eqref{post_correction_known_biases} and \Eqref{second_known_bias_full}. \Figref{bias_correction_illustration_1e4} shows the results when using calibration sets of $10^4$ mock shear values, and \Figref{bias_correction_illustration_1e6} shows the results for $10^6$ mock shear values. The values plotted in these figures are detailed in \Appref{supp_data}.

As we find that $c$ is corrected perfectly on average even in the case of the smaller calibration set with a first-order correction, and the scatter in actual $c$ values after bias correction is identical to the scatter in measured $c$ values prior to correction, we do not show it in these plots.

Consistent with our predictions, the first-order correction for $m$ brings the mean residual bias closer to zero, but doesn't correct it perfectly. The mean distance from zero is, however, much smaller than the scatter. The second-order correction is superior at this, but still not perfect. The scatters in both corrections are consistent with our predicted scaling with $m$, and we see that the scatters of post-correction biases increase when the second-order correction is applied.

Notably, when dealing with cases of sub-per-cent pre-calibration biases, we observe that the calibration procedure can actually do more harm than good if the size of the calibration set is too small. The $N=10^6$ size calibration set we use here results in the scatter of post-calibration $m$ values being of order $1$ per-cent, which implies that if a shear estimation already has a bias of this order or less, calibration with this size of a dataset is likely to result in a stronger bias than if no calibration had been performed. This effect can be mitigated through the use of a larger calibration set, but there will always exist a bias threshold below which calibration will be counter-productive. One might thus consider only calibrating if the bias is measured to be above a certain threshold; we discuss this possibility in \Secref{cond_calibration}.

\begin{figure*}
	\includegraphics[scale=0.54]{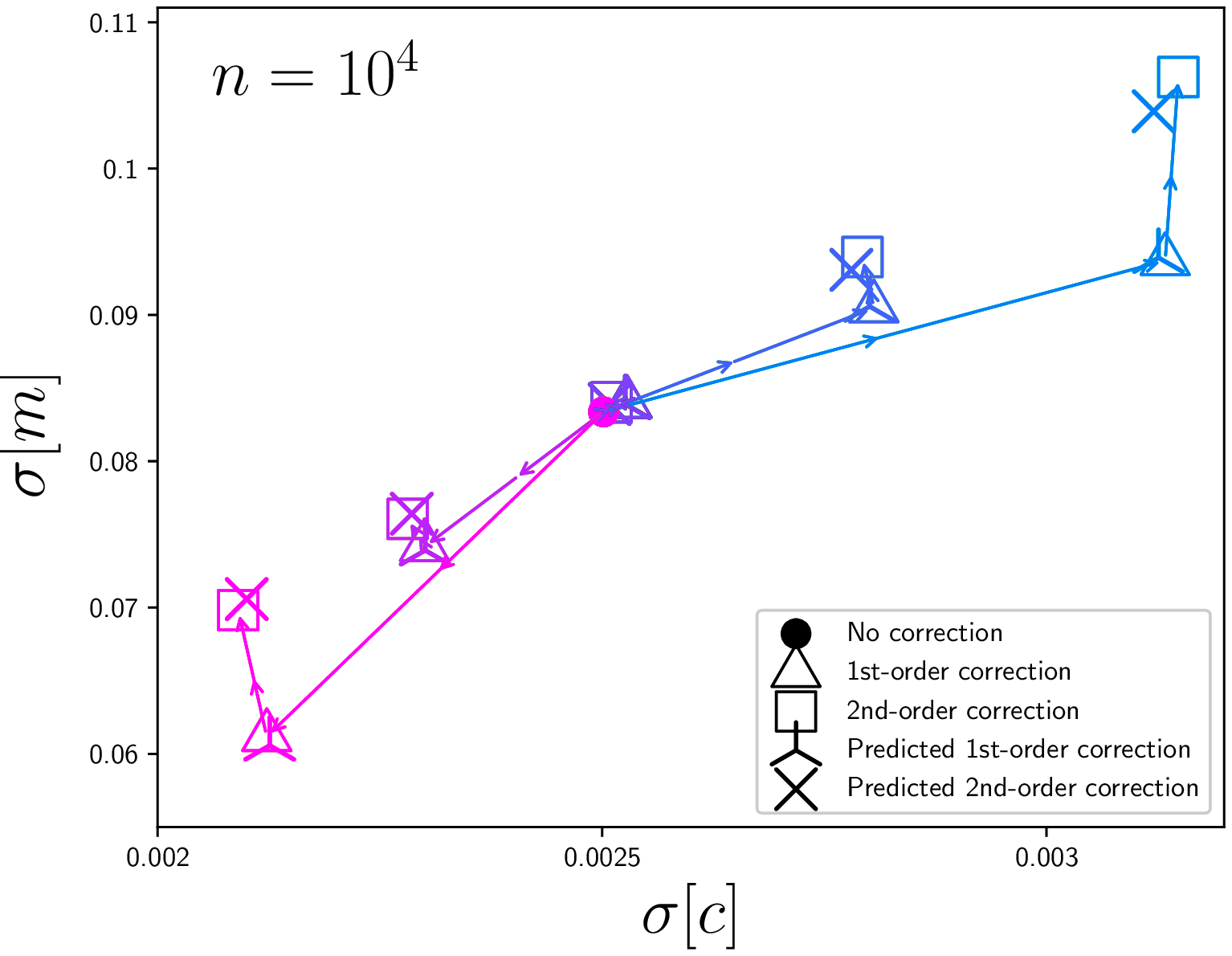}
	\includegraphics[scale=0.54]{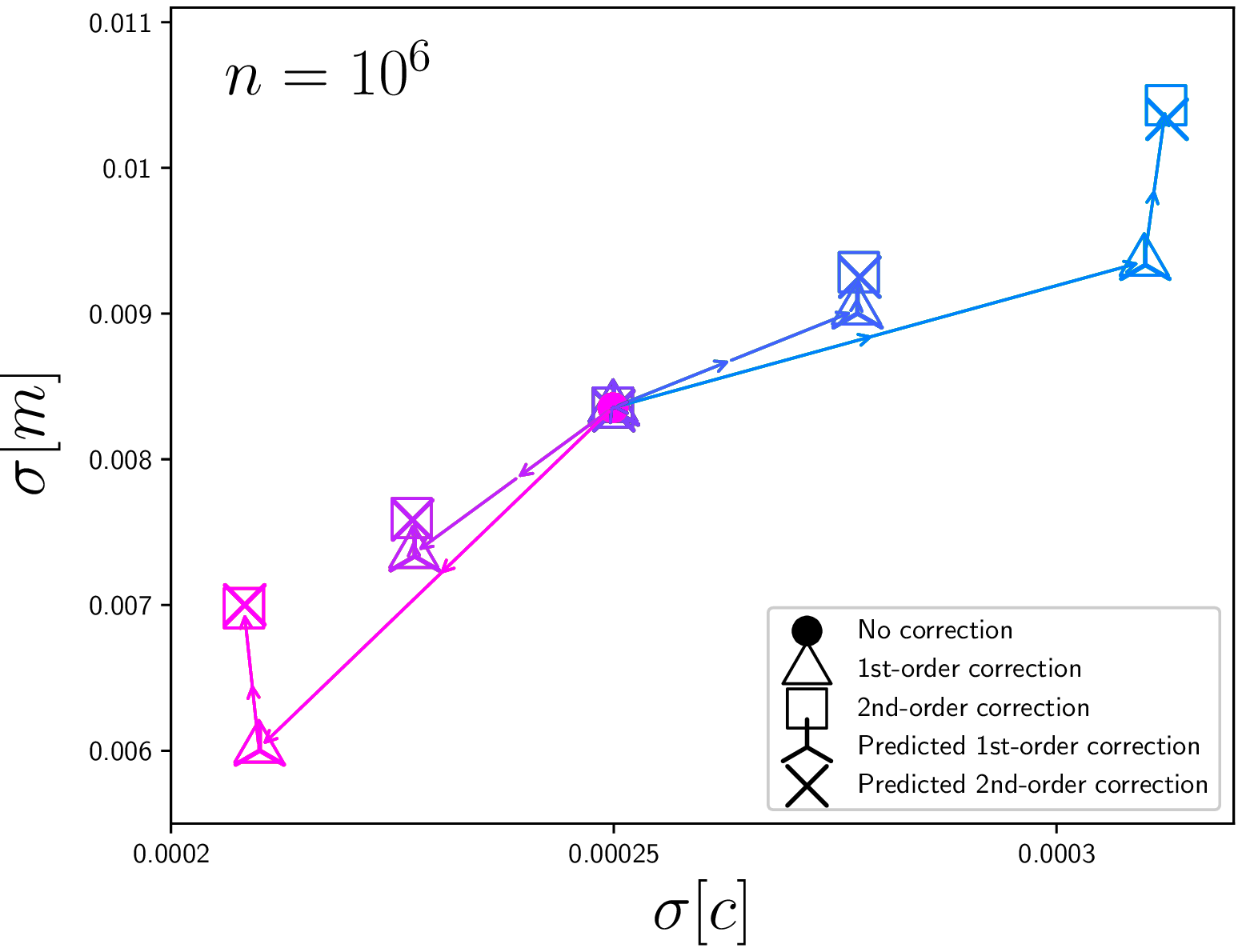}
	\caption{An illustration of the effects of calibration on the scatters of the $m$ and $c$ bias parameters. Solid markers show the scatter in the measured bias parameters before correction, triangles/squares show the scatter in the actual bias parameters after a first/second-order correction, and inverted triangles/squares show our predictions for it after a first/second-order correction using. Arrows are drawn connecting points for successive orders of calibration. The left figure uses realisations of simulations of $10^4$ mock shear values, and the right figure uses realisations of simulations of $10^6$ mock shear values. The colours correspond to the $m$ values in the rightmost column of \Figref{bias_correction_illustration_1e4}: magenta for $m=0.2$ (lower-left-most set of markers and arrows), purple for $m=0$ (centre), cyan for $m=-0.2$ (upper-right-most), and intermediate shades for $m=-0.1$ and $m=0.1$. The values for $m=-0.01$ and $m=0.01$ are not shown here since there's no observable deviation from a linear relationship in this regime. }
	\Figlabel{bias_correction_scatter_illustration}
\end{figure*}

\Figref{bias_correction_scatter_illustration} shows the change in the scatter of the bias parameters for the first- and second-order corrections, along with our predictions for the scatter after the first-order correction, as given in \Eqref{post_correction_vars}. This shows that the changes in scatter for the first-order correction are consistent with our predictions, and we can also see that the second-order correction increases the scatter on the residual $m$ bias, as expected.

\begin{figure*}
	\includegraphics[scale=0.54]{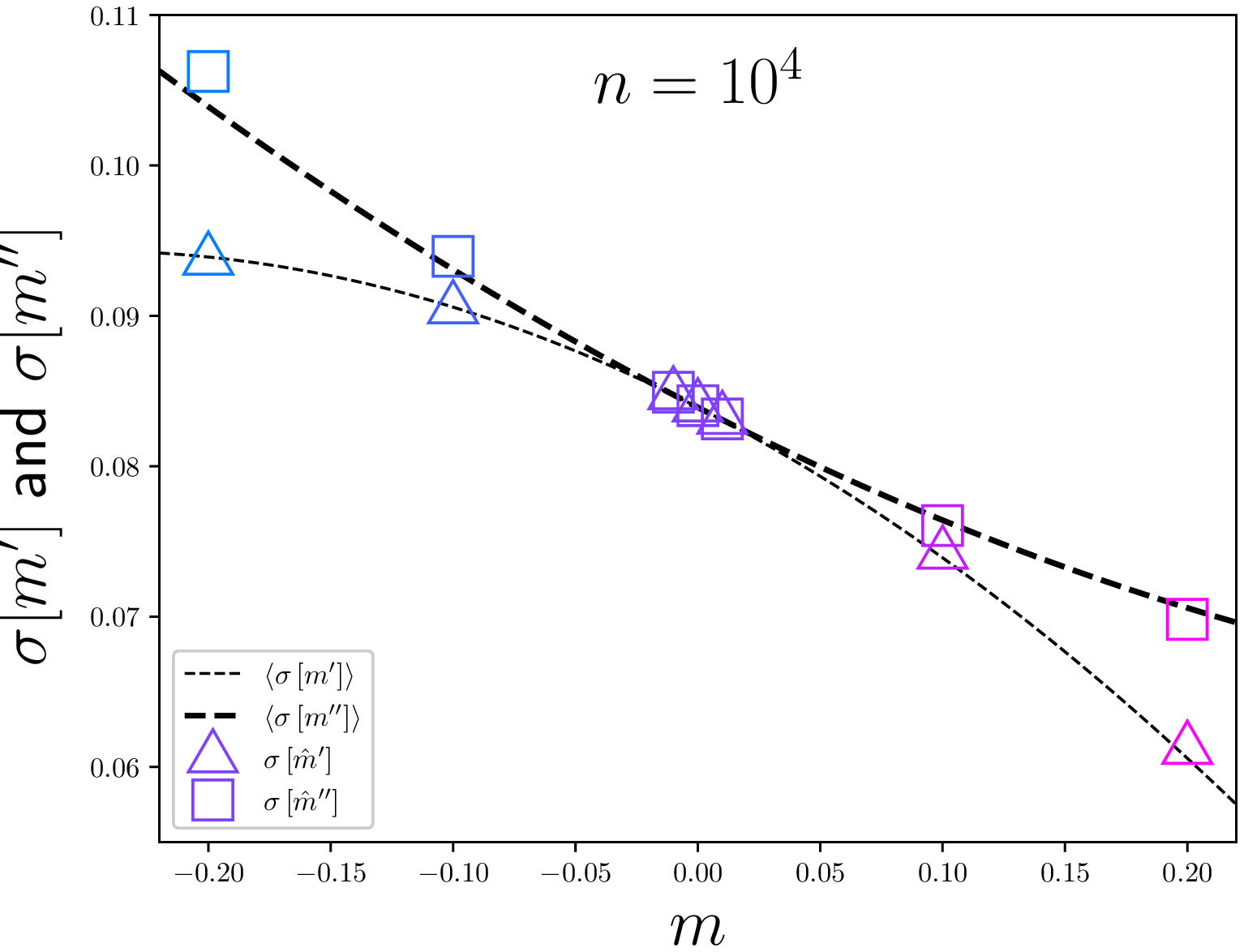}
	\includegraphics[scale=0.54]{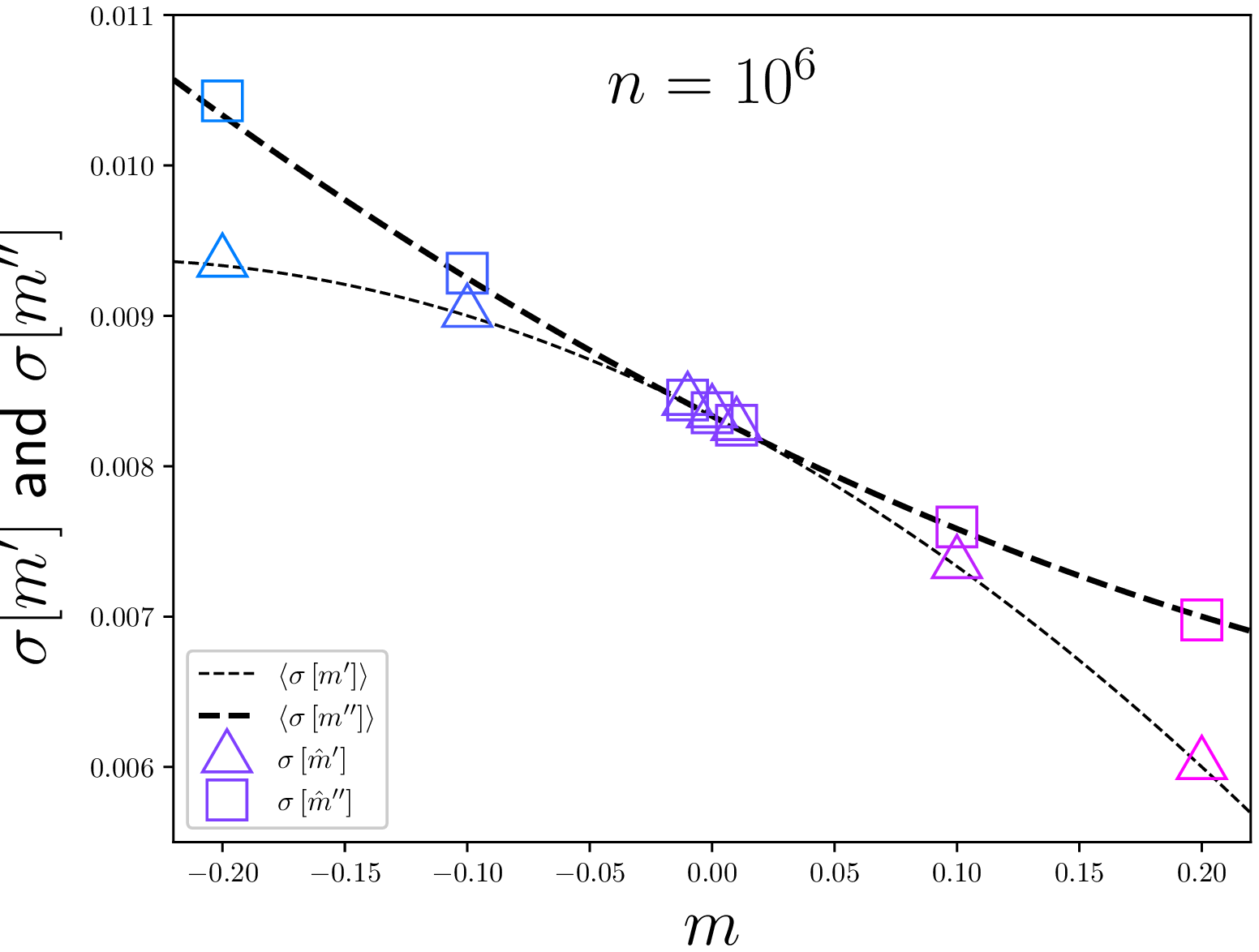}
	\caption{The residual $m$ bias and scatter after first- and second-order corrections plotted against the initial multiplicative bias $m$, for an $n=10^4$ simulation (left) and an $n=10^6$ simulation (right). Lines show predictions using our calculations, and points show the mean values measured from our mock calibrations. Errors on the plotted points are smaller than the marker sizes.}
	\Figlabel{bias_correction_proj_m_scatter_illustration}
\end{figure*}

\Figref{bias_correction_proj_m_scatter_illustration} shows the dependence of the residual $m$ and its scatter on the initial $m$, from our calculations in \Eqref{post_correction_known_biases} as well as the results of the simulations. This shows that there is indeed a moderate dependence, with larger $m$ values corresponding to more positive (and generally larger) residual $m$ biases after the first-order correction, as well as corresponding to smaller scatters on this residual bias. There is very good agreement between our predictions and the results of the simulations for both the residual $m$ and its scatter.

As the increase in the scatter of the residual $m$ bias for the second-order correction is of comparable magnitude to the change in the expected post-correction value, it is worth considering if this correction is overall beneficial. To make this judgment, we require some manner of quantifying the quality of a calibration, taking into account the expected value and its scatter. This may vary depending on the purpose of this calibration, but one reasonable choice is to use the root-mean-square distances of the post-calibration $m$ and $c$ values from zero:
\begin{align}
	\Eqlabel{dmc_definitions}
	d_m &= \sqrt{ \expec{m}^2 + \var{m} } \mathrm{,} \\ \nonumber
	d_c &= \sqrt{ \expec{c}^2 + \var{c} } \mathrm{.}
\end{align}
These can be combined into an overall distance parameter $d$ by scaling these distances by the target values of $m$ and $c$, which will vary by application:
\begin{equation}
	\Eqlabel{d_definition}
	d = \sqrt{\left(\frac{d_m}{m_t}\right)^2 + \left(\frac{d_c}{c_t}\right)^2} \mathrm{.}
\end{equation}
For this paper, we will use the target values used by the Euclid mission \citep{LauAmiArd11}:
\begin{align*}
	m_t &= 2 \times 10^{-3}\mathrm{,} \\
	c_t &= 5 \times 10^{-5}\mathrm{.}
\end{align*}
A value of $d<1$ will thus indicate that the root-mean-square $m$ and $c$ are both less than their target values, and $(m,c)$ lies within the ellipse centred at $(0,0)$ with axes lengths $m_t$ and $c_t$.

\begin{figure*}
	\includegraphics[scale=0.54]{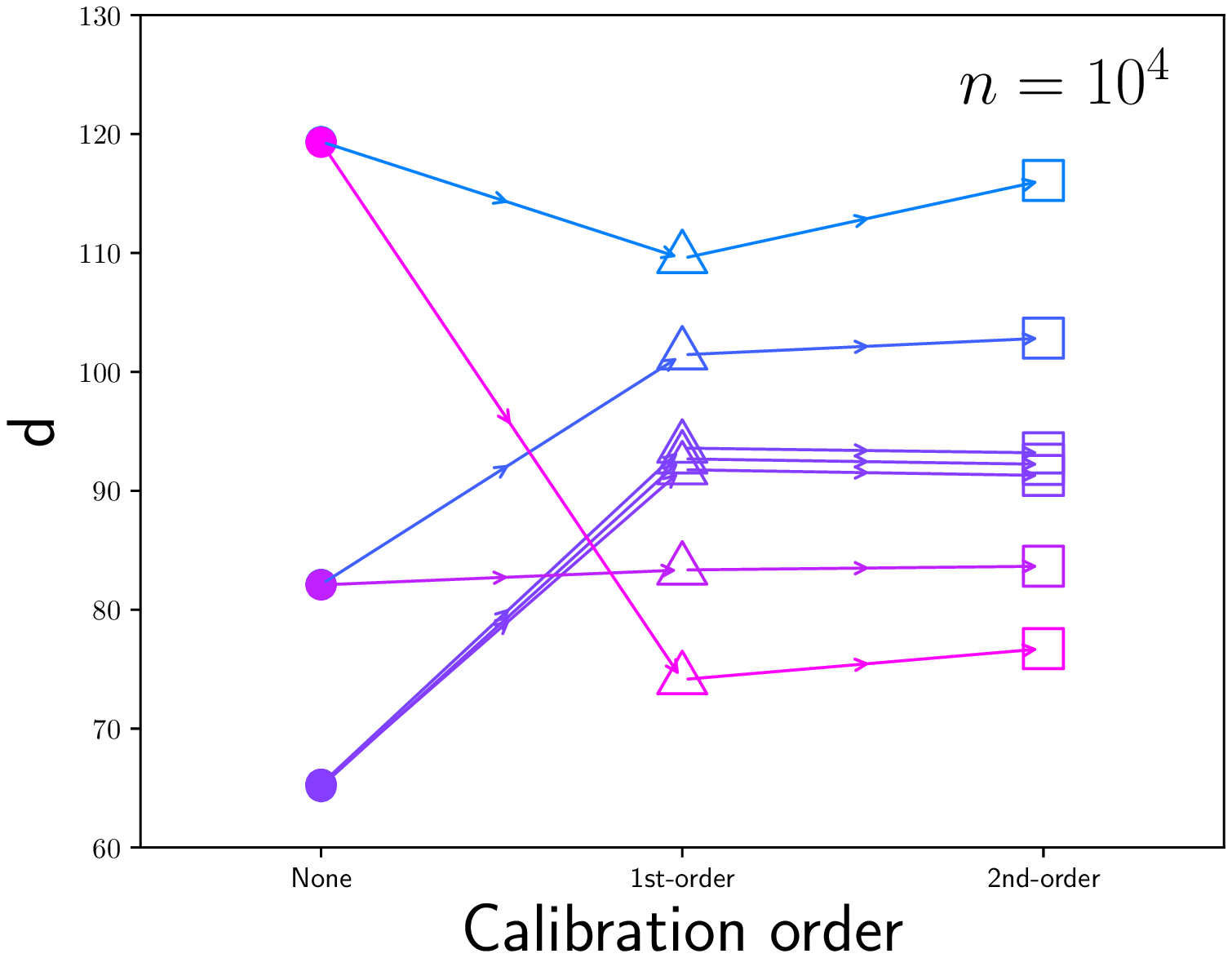}
	\includegraphics[scale=0.54]{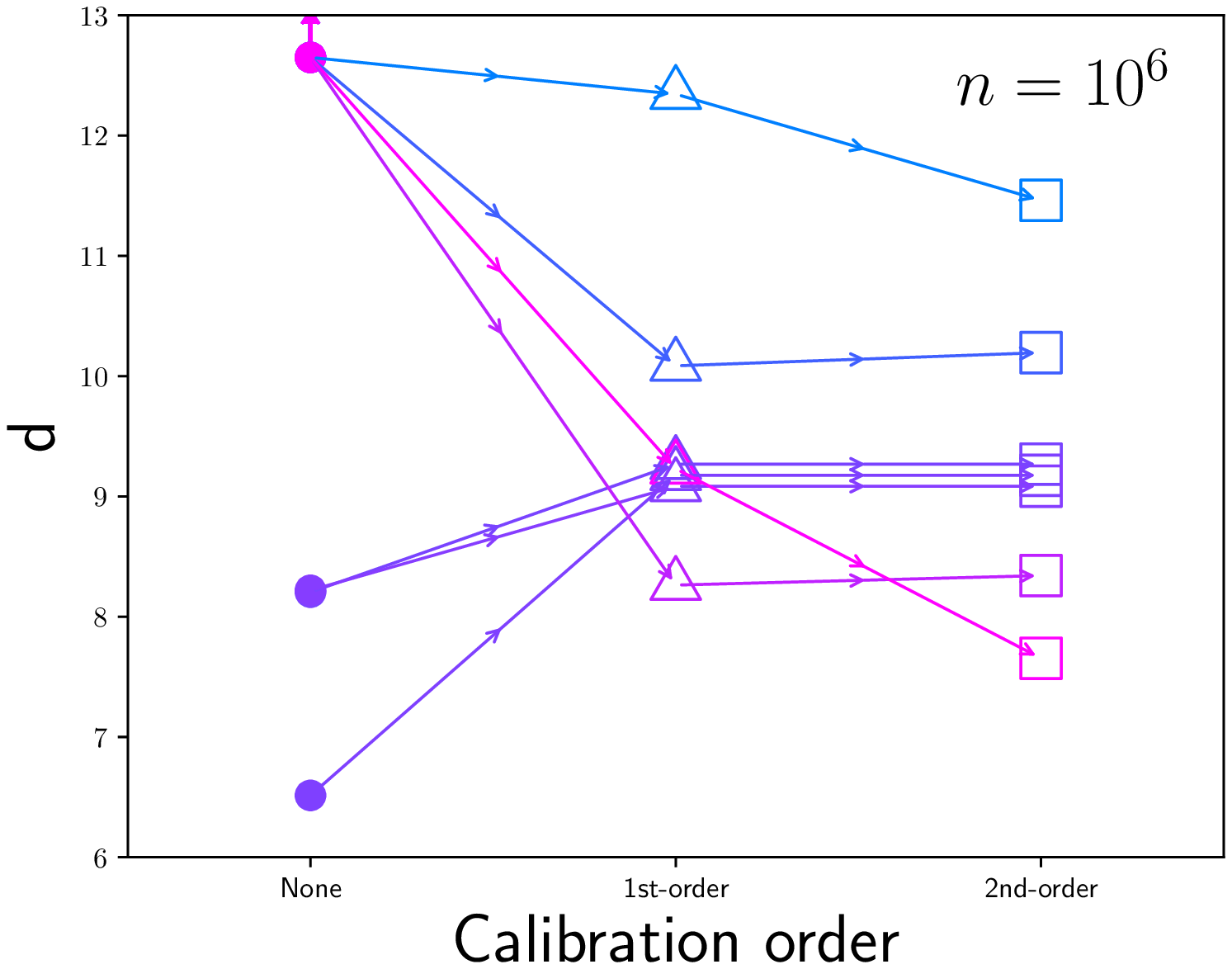}
	\caption{The calibration quality parameter $d$, as defined in \Eqref{d_definition}, for both sizes of datasets and for both orders of calibration. Points correspond to different actual $m$ biases, with the same colours used as in \Figref{bias_correction_illustration_1e4}. We use $c=0$ for all points plotted here, as the perfect correction of $c$ bias results in there being no difference in the quality of the correction aside from in the uncorrected case. Arrows are shown connecting points for successive calibrations to aid viewing. Note that some points for the no-calibration scenario lie far above the upper bound of the plot, indicated by the points with upward-pointing arrows.}
	\Figlabel{bias_correction_quality_illustration}
\end{figure*}

We show the values of this quality parameter for both sizes of datasets and both calibration orders in \Figref{bias_correction_quality_illustration}. This shows that the first-order calibration is highly beneficial except in the $m=0,\;c=0$ scenario. The second-order correction provides no advantage over the first-order correction in the $\left|m\right| \leq 0.1$ scenarios, but does provide a significant benefit with the larger simulation size when $m=-0.2$ or $0.2$. This implies that the optimal correction will depend on the initial $m$, with larger $m$ requiring a higher-order calibration method.

The case of $m=0,\;c=0$ requires some discussion, since it appears that calibration is counter-productive in this case. This is a true: If a shear estimation method is perfectly unbiased, any attempt to calibrate it can only do nothing or induce bias. However, one can only know that a method is perfectly unbiased through measuring the bias, and this measurement can then be used to correct for it. One might consider not applying a calibration if the measured bias is sufficiently low, however. We discuss this option in \Secref{cond_calibration}.

\section{Predictions}
\Seclabel{projections}

Let us estimate what the magnitude of the residual biases will be for a calibration set of a given size. We will first assume that the noise in shear measurements is independent of the shear, with the shear-measurement algorithm providing a noise-free but biased estimate of the shape, and that the measurement noise is normally-distributed.  We will also assume that the shear values used in the calibration set are normally distributed with standard deviation $\sigma_{\g}$, and that the shear is small enough that shear values can be added linearly. For example, a given galaxy will be first assigned a shear $\g_i$ and measurement noise $\delta_{\g,i}$. The measured shape will then be $\hat{\g_i} = (1+m)\g_i+c+\delta_{\g,i}$.

We will only make predictions using a first-order calibration here, but the same logic and similar calculations can be used for higher-order calibrations.

The bias components can be measured through a linear regression of the measured shapes against the input shears. Since we are assuming that the measurement error is the same for all galaxies, this can be done through a standard least-squares regression. In this case, the standard errors on the measured $m$ and $c$ will be
\begin{align}
	\std{m} &= \sqrt{ \frac{ \frac{1}{n-2} \sum^n_{i=1} \delta_i^2 }{ \sum^n_{i=1}(\g_{s,i}-\overline{\g_s})^2 } } \mathrm{,} \\ \nonumber
	\std{c} &= \std{\hat{m}}\sqrt{ \frac{1}{n}\sum^n_{i=1} \g_{s,i}^2 } \mathrm{,}
\end{align}
where $\delta_i = \g_{m,i} - [(1+\hat{m})\g_i+\hat{c}]$ is the deviate from the best fit and $n$ is the number of galaxies in the calibration set. In the large-$n$ limit, this converges to:
\begin{align}
	\Eqlabel{stddev_projections}
	\std{m} &= \frac{1}{\sqrt{n}}\frac{ \std{\g} }{ \sigma_{\g} } \\ \nonumber
	\std{c} &= \frac{ \std{\g} }{ \sqrt{n} } \mathrm{.}
\end{align}
If the correction proposed here is applied, the mean residual biases can be calculated from \Eqref{post_correction_known_biases} to be
\begin{align}
	\expec{m'} &= \frac{1}{n}\frac{ \var{\g} }{ \sigma_{\g}^2 } (1 + m) + m^3 \mathrm{,} \\ \nonumber
	\expec{c'} &= 0 \mathrm{,}
\end{align}
and the new errors can be estimated from \Eqref{post_correction_vars}:
\begin{align}
	\std{m'} &\approx \frac{1}{\sqrt{n}}\frac{ \std{\g} }{ \sigma_{\g} } \left[1 - m\right]  \\ \nonumber
	\std{c'} &\approx \frac{ \std{\g} }{ \sqrt{n} } \left[1 - m\right] \mathrm{.}
\end{align}
We see here that the mean residual bias in $m$ scales as $1/n$ if the $m^3$ term is disregarded, while the error on it scales as $1/\sqrt{n}$. This implies that, unless $m$ is large enough for the $m^3$ term to become significant, the known component of the bias will generally be negligible compared to the unknown component.

Note that these predictions are made using the actual value of $m$. In practice, only the measured value $\hat{m}$ will be known, and so it must be used instead, which will result in a slight bias in these projections. We can reduce this bias through a correction term which we find by replacing $m$ in the above equations with $\hat{m} = m + \delta_m$ and taking the expectation value. For example, for $\expec{m'}$ we get:
\begin{align}
	\expec{m'} - m'_c &= \frac{1}{n}\frac{ \var{\g} }{ \sigma_{\g}^2 } (1 + \expec{m + \delta_m}) + \expec{(m + \delta_m)^3} \\ \nonumber
		&= \expec{m'} + 3m\var{m} \\ \nonumber
	m'_c &= -3m\var{m} \mathrm{.}
\end{align}
This correction term also depends on $m$, but as $\hat{m}$ is an unbiased estimate of $m$, we can use it here without need for further calculations. This gives us the unbiased estimate of the known $m$ bias component:
\begin{equation}
	\Eqlabel{post_correction_known_biases_estimate}
	\expec{m'} = \frac{1}{n}\frac{ \var{\g} }{ \sigma_{\g}^2 } (1 - 2\hat{m}) + \hat{m}^3 \mathrm{.}
\end{equation}
For $c$, the correction is already unbiased, and so no correction is necessary.

This procedure can be applied to the post-correction variance as well. The unbiased variance estimates can be calculated to be:
\begin{align}
	\Eqlabel{post_correction_known_vars_estimate}
	\var{m'} &= \var{m}\big( 1 - 2\hat{m} - 3\hat{m}^2 + 5\var{m} + 4\hat{m}^3 \\ \nonumber
	         &\;\;\;\;- 8\hat{m}\var{m} + 4\hat{m}^4 - 22\hat{m}^2\var{m} + 18\sigma^4\left[m\right] \big) \mathrm{,} \\ \nonumber
	\var{c'} &= \var{c}\big( 1 - 2\hat{m} + 3\hat{m}^2 - 2\hat{m}^3 + \hat{m}^4 + 2\sigma^4\left[m\right] \big)\mathrm{,}
\end{align}
where $\std{m}$ and $\std{c}$ are defined as in \Eqref{stddev_projections}. Note that the higher-order terms here were calculated based on an assumption of Gaussianity in the distribution of $\hat{m}$, which is not expected to be the case in reality, and so this should be treated as an approximation only.

\begin{figure}
	\includegraphics[scale=0.45]{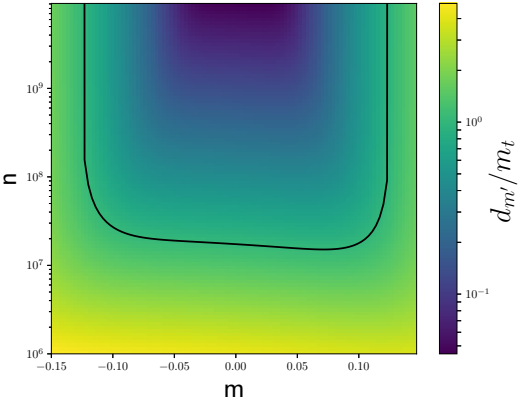}
	\caption{The projected $d_{m'}$ as a function of initial multiplicative bias $m$ and number of simulated galaxies in the calibration set $n$. The black line indicates $d_{m'}=m_t$, using the target for the Euclid mission $m_t=0.002$; above this line, the root-mean-square post-calibration $m$ bias will be less than the target value.}
	\Figlabel{d_projections}
\end{figure}

This analysis can also be used to predict the required size of a calibration set. When $d_m,d_c < m_t,c_t$, the typical bias parameters will be within the target values. As $d_m$ is typically dominant, we will focus on it. Expanding the definition of $d_m$ given in \Eqref{dmc_definitions} using the projections for $\expec{m'}$ and $\var{m'}$, we can calculate the projected $d_{m'}$ to be:
\begin{align}
	d_{m'} &= \sqrt{\bigg( 3\sigma^4\left[m\right](1+2m+m^2)} \\ \nonumber
	       &\;\;\;\;\overline{+ \var{m}(1 - 2m - 3m^2 + 5m^3 + 5m^4) + m^6\bigg)} \mathrm{.}
\end{align}
Using \Eqref{stddev_projections}, we can then express this as:
\begin{align}
	d_{m'} &= \sqrt{ 3\frac{1}{n^2}\frac{ \sigma^4\left[\g\right] }{ \sigma^4_{\g} }(1+2m+m^2)} \\ \nonumber
	       &\;\;\;\;\overline{+ \frac{1}{n}\frac{ \var{\g} }{ \sigma_{\g}^2 }(1 - 2m - 3m^2 + 5m^3 + 5m^4) + m^6} \mathrm{.}
\end{align}
To project the needed number of simulations, we set $d_{m'}=m_t$ and solve for $n$ to get
\begin{align}
	\Eqlabel{n_projection}
	n &=  \frac{ \var{\g} }{ \sigma_{\g}^2 } \frac{1}{m_t^2 - m^6} \left( 1 - 2m - 3m^2 + O(m^3) \right) 
\end{align}
This relationship can be seen in \Figref{d_projections}. Note that this gives $n$ for a best-case scenario, where the calibration dataset is expected to be identical to the dataset on which a measurement will be made. This is not likely to be the case in practice, and handling this fact will result in the required number of simulations increasing proportional to the number of bins ($n\sbr{b}$) required to represent the full range of possible complicating parameters. This will likely be at least of order $10$, and possibly much higher.

As it is impossible to know $m$ before performing running a simulation, $n$ must be estimated based on the expected worst-case scenario, which will come from the most negative value of $m$ which is reasonably expected to occur. For a value of $m=-0.1$ and using the other simulation parameters defined in \Secref{illustration}, this gives $n=2.7 \times 10^7 n\sbr{b}$, and for a value of $m=-0.01$, this gives $n=1.8 \times 10^7 n\sbr{b}$. 

Using the fiducial values for our simulations and the assumption that $|m|<0.1$, we can express the scaling of $n$ from \Eqref{n_projection} as
\begin{equation}
	n \approx 2.7 \times 10^7 n\sbr{b} \left(\frac{2\times10^{-3}}{m_t} \cdotp \frac{ \std{\g} }{ 0.25 } \cdotp \frac{ 0.03 }{ \sigma_{\g} }\right)^2 \mathrm{.}
	\Eqlabel{n_sims_estimate}
\end{equation}
For instance, a simulation with $\std{g_s}=0.01$ instead of the $0.03$ used here will require approximately $9$ times as many galaxies, for a size of $n \approx 2.4 \times 10^8 n\sbr{b}$.

Note that \Eqref{n_projection} breaks down if $\left|m^3\right| \ge m_t$. This is because the first-order calibration has an expected residual multiplicative bias with an $m^3$ term which does not scale with $n$. If it is expected that $m$ might be this large, then the second- (or even higher) order calibration will be necessary. Although this has the drawback of increasing the scatter on the post-correction multiplicative bias, this scatter will still scale with $n$, and so it will be possible to reduce it sufficiently with a large enough set of simulated data.

\section{Alternative Methods}
\Seclabel{alt_methods}

In this paper, we have discussed the bias correction that is necessary for the case of a frequentist method which is discovered to have a bias after comparison to a calibration dataset. It is worth considering other possible shear-measurement scenarios. While it is impossible to envision all possibilities, we will discuss two general classes of alternative methods in this section to provide a framework for future analyses. In \Secref{machine_learning}, we discuss machine-learning methods, where some parameters of the method are necessarily determined through calibration to some dataset, and in \Secref{bayesian_methods}, we discuss Bayesian approaches to accounting for bias. In \Secref{cond_calibration}, we additionally discuss whether it might be beneficial to calibrate conditionally based on the measured bias.

\subsection{Machine-Learning Methods}
\Seclabel{machine_learning}

One approach to shear-measurement involves designing a method with a number of a free parameters which are determined through a comparison to a calibration dataset. One example of this is MegaLUT \citep{TewCanCou12}, which uses the calibration dataset to build a large lookup table which maps the galaxy elongation, the point-spread-function (PSF) elongation, the size ratio of the galaxy and PSF, and the relative orientation of the galaxy and PSF to the proper correction to the galaxy's observed ellipticity.\footnote{Note that MegaLUT is specifically a method for estimating galaxy ellipticity before the influence of the PSF, not shear. It is nevertheless a useful comparison, as to first order, this ellipticity will be the galaxy's shear plus a random component due to shape noise.} It is thus effectively a piecewise-defined function of these parameters, with the parameters characterizing each piece determined through the calibration step. In general, all machine-learning methods can be considered to provide a calibration-determined function of some observable parameters.

We can consider the process of gravitational shear acting on the undistorted images of galaxies to produce the observed images as a functional operation, albeit a non-deterministic one due to various random factors such as the undistorted galaxy ellipticity and noise in observations. Let us express this as
\begin{equation}
	\mathbf{F}(\g_1,\g_2,x_1,...,x_n) = (y_1, ..., y_n)\mathrm{,}
\end{equation}
where $\g_1$ and $\g_2$ are the two shear components, $x_1$ through $x_n$ are other properties of the galaxy and its observation (eg. its shape, pixel noise, etc.), and $y_1$ through $y_n$ are the observed properties of the galaxy. The goal of a machine-learning method is to generate a function $G(y_1,y_2,...,y_n)$ such that
\begin{equation}
	\Eqlabel{expec_G_of_F}
	\expec{\mathbf{G}\big(\mathbf{F}(\g_1,\g_2,x_1,x_2,...,x_n)\big)} = (\g_1, \g_2) \mathrm{.}
\end{equation}
The definition of the expectation value here must be clarified. Ideally, $G$ will provide unbiased estimates of $\g_1$ and $\g_2$ for any possible non-noise parameters, requiring only the noise parameters to be marginalized over. At minimum, it should provide unbiased shear estimates if all parameters are marginalized over.

Let us work through an example with a toy model, simplifying to the case of a single shear parameter $\g$. We will assume that $F$ takes the form
\begin{equation}
	F\sbr{toy}(\g,\delta_{\g}) = \hat{\g} = (1+m)\g+c + \delta_{\g}\mathrm{,}
\end{equation}
where $\delta_{\g}$ is a noise parameter and $m$ and $c$ are the multiplicative and additive bias components, as before. We will also assume that $G$ takes the form
\begin{equation}
	G\sbr{toy}(\hat{\g}) = \hat{g}' = a \hat{\g} + b\mathrm{,}
\end{equation}
where $a$ and $b$ are parameters which are fit based on the calibration data. We desire to satisfy \Eqref{expec_G_of_F}, which gives us the constraint
\begin{equation}
	\expec{ a \hat{\g} + b } = \g \mathrm{.}
\end{equation}
We can accomplish this by ensuring that it is true on average for the calibration set:
\begin{align}
	\frac{1}{n} \sum_{i=0}^n \left(a \hat{\g_i} + b - \g_i \right) &= 0\mathrm{,}
\end{align}
but this constraint alone isn't enough to uniquely define both $a$ and $b$. A natural way to do this is through minimizing the mean square difference between $a \hat{\g_i} + b$ and $\g_i$. Alternatively, we could minimize the mean square difference in the inverse operation, between $\hat{\g_i}$ and $(\g_i-b)/a$. These approaches will give distinct, but related, results. In this case, $\g_i$ are known exactly, and we expect there to be measurement error on $\hat{\g_i}$, so it makes the most sense to minimize the latter comparison.

At this point, it is apparent that we are performing the same procedure as in \Secref{projections}, performing a linear regression of $\hat{\g} = (1+m)\g + c$ where $1/a$ corresponds to $1+m$ and $-b/a$ corresponds to $c$. In the case of this toy model at least, machine-learning is no different from calibration, and we can thus expect it to face the same issues with post-calibration bias.

It is worth discussing other approaches that might be taken here. One approach would be to instead attempt a linear regression of $\g = a\hat{\g}+b$, in the hope that the estimate of $a$ would provide us with an unbiased estimate of $1/(1+m)$. Unfortunately though, this approach does not work. The reason for this can be seen in the relationship between the slopes of linear regressions of $y$ versus $x$ and $x$ versus $y$:
\begin{equation}
	a_{yx} = \frac{\var{y}}{\var{x}} a_{xy} \mathrm{,}
\end{equation}
where $a_{yx}$ is the fitted slope for a linear regression of $y$ versus $x$ and $a_{xy}$ is the same for $x$ versus $y$, and $\var{x}$ and $\var{y}$ are the variances of the $x$ and $y$ points being fit, respectively. In this case, we have
\begin{align}
	\var{\g} &= \sigma_{\g}^2 \\ \nonumber
	\var{\hat{\g}} &= \sigma_{\hat{g}}^2 + (1+m)^2\sigma_{\g}^2 \mathrm{,}
\end{align}
which gives us the relationship
\begin{equation}
	a_{\g \hat{\g}} = \frac{\sigma_{\g}^2}{\sigma_{\hat{g}}^2 + (1+m)^2\sigma_{\g}^2} a_{\hat{\g} \g} \mathrm{.}
\end{equation}
This means that if $a_{\hat{\g} \g}$ is an unbiased estimate of $1+m$, then
\begin{equation}
	\expec{a_{\g \hat{\g}}} = \frac{\sigma_{\g}^2}{\sigma_{\hat{g}}^2 + (1+m)^2\sigma_{\g}^2} (1+m) \mathrm{.}
\end{equation}
This would only give us an unbiased estimate of $1/(1+m)$ if $\sigma_{\g}^2 \gg \sigma_{\hat{\g}}^2$. In practice, we expect the opposite to generally be true, and so unfortunately this approach cannot be used to determine an unbiased correction.

It is impossible for us to test every conceivable machine-learning method, but the analysis of our toy model here makes it appear likely that most such methods will have similar issues with post-calibration bias. It will be necessary to test any proposed method on multiple calibration datasets to determine if this is indeed the case, and to quantify the expected post-calibration bias.

\subsection{Bayesian Methodology}
\Seclabel{bayesian_methods}

The approach taken to this point has been purely frequentist, using only a single estimator for $\g$. An alternative to this is to take a Bayesian approach and construct a likelihood function for $\g$ given $\hat{\g}$, $\hat{m}$, and $\hat{c}$. Using Bayes' Theorem, we have
\begin{equation}
	P(\g|\hat{\g},\hat{m},\hat{c}) = \frac{P(\hat{\g}|\g,\hat{m},\hat{c})P(\g)}{P(\hat{\g})}\mathrm{.}
\end{equation}
$P(\g)$ represents the prior knowledge on the likelihood of $\g$, and so is beyond the scope of this work, and $P(\hat{\g})$ can be determined through normalizing such that the likelihood sums to $1$. We will thus only need to determine $P(\hat{\g}|\g,\hat{m},\hat{c})$. Let us start by assuming the likelihoods
\begin{align}
	P(\hat{g}|g,m,c) &= \exp\left[ -\frac{1}{2\var{g}} \left((1+m)\g + c - \hat{\g} \right)^2 \right] \\ \nonumber
	P(m|\hat{m}) &= \exp\left[ -\frac{1}{2\var{m}} \left(m - \hat{m} \right)^2 \right] \\ \nonumber
	P(c|\hat{c}) &= \exp\left[ -\frac{1}{2\var{c}} \left(c - \hat{c} \right)^2 \right] \mathrm{.}
\end{align}
We can then determine $P(\hat{\g}|\g,\hat{m},\hat{c})$ by marginalizing $P(\hat{g}|g,m,c)$ over $m$ and $c$:
\begin{align}
	\Eqlabel{likelihood_P_ghat}
	P(\hat{\g}|\g,\hat{m},\hat{c}) &= \int_{-\infty}^{\infty} \int_{-\infty}^{\infty} P(\hat{g}|g,m,c) P(m|\hat{m}) P(c|\hat{c}) dm dc \\ \nonumber
		&= A \frac{\exp\left[ -\frac{1}{2}\frac{\left((1+\hat{m})\g + \hat{c} - \hat{\g} \right)^2}{g^2\var{m}+\var{c}+\var{g}} \right] }{\sqrt{g^2\var{m}+\var{c}+\var{g}}} \mathrm{,}
\end{align}
where $A$ is a normalization factor such that $P(\hat{\g}|\g,\hat{m},\hat{c})$ sums to $1$.

If we wish to use a single estimator, we could take either the mean of \Eqref{likelihood_P_ghat} or its maximum likelihood. Unfortunately, there is no analytic solution for its mean, and the solution for its maximum likelihood is extremely complicated. To analyse this, we will thus use a Gaussian approximation. We expect $\var{c}+\var{g} \gg g^2\var{m}$, and so we can approximate based on this assumption by rewriting \Eqref{likelihood_P_ghat} as
\begin{align}
	\Eqlabel{likelihood_P_ghat_rewritten}
	&P(\hat{\g}|\g,\hat{m},\hat{c}) = \\ \nonumber
	&\;\;\;\;A \exp\left[ -\frac{1}{2(\var{c}+\var{g})}\frac{\left((1+\hat{m})\g + \hat{c} - \hat{\g} \right)^2}{1+\frac{g^2\var{m}}{\var{c}+\var{g}}} \right]  \\ \nonumber
	&\;\;\;\;\times\left(1 + \frac{g^2\var{m}}{\var{c}+\var{g}}\right)^{-1/2} \mathrm{.}
\end{align}
Note here that the zeroth-order approximation to this will simply be a Gaussian centred around $\g = (\hat{\g}-\hat{c})/(1+\hat{m})$, which is in fact the form of the intuitive bias correction we presented in \Eqref{intuitive_bias_correction}. The first-order approximation can be found in \Appref{exp_calcs}, and it gives us the maximum likelihood estimate for $\g$,
\begin{equation}
	\Eqlabel{bayesian_g_estimate}
	\hat{\g}^{\rm b} = \frac{(\hat{\g} - \hat{c})(1+\hat{m})}{(1+\hat{m})^2 + \var{m} - (\hat{\g}-\hat{c})^2 \var{m}/(\var{c} + \var{g})} \mathrm{,}
\end{equation}
and its expected variance,
\begin{equation}
	\Eqlabel{bayesian_g_variance}
	\var{\hat{\g}^{\rm b}} = \frac{\var{c}+\var{g}}{(1+\hat{m})^2 + \var{m} - (\hat{\g} - \hat{c})^2 \frac{\var{m}}{\var{c} + \var{g}}} \mathrm{.}
\end{equation}
This can be approximated as
\begin{align}
	\Eqlabel{bayesian_g_variance_approx}
	\var{\hat{\g}^{\rm b}} &\approx \frac{\var{c}+\var{g}}{(1+\hat{m})^2 + \var{m}} + \frac{(\hat{\g} - \hat{c})^2\var{m}}{((1+\hat{m})^2 + \var{m})^2}\mathrm{,}
\end{align}
allowing us to estimate the three variance terms to be
\begin{align}
	\Eqlabel{bayesian_g_variance_components_approx}
	\var{m^{\rm b}}  &= \frac{\var{m}}{(1+\hat{m})^2 + \var{m}} \mathrm{,}\\ \nonumber
	\var{c^{\rm b}}  &= \frac{\var{c}}{(1+\hat{m})^2 + \var{m}} \mathrm{, and}\\ \nonumber
	\var{\g^{\rm b}} &= \frac{\var{\g}}{(1+\hat{m})^2 + \var{m}} \mathrm{.}
\end{align}
Note here that $\var{m^{\rm b}}$ is more of an approximation than the other terms, as it properly would require us to write \Eqref{bayesian_g_variance_approx} in terms of $\g$ rather than $\hat{\g}$. These variances are in fact substantially similar to those for our frequentist estimator, as presented in \Eqref{post_correction_vars}, with the same first-order dependence on $m$ (although here expressed in terms of $\hat{m}$ instead).

\begin{figure*}
	\includegraphics[scale=0.53]{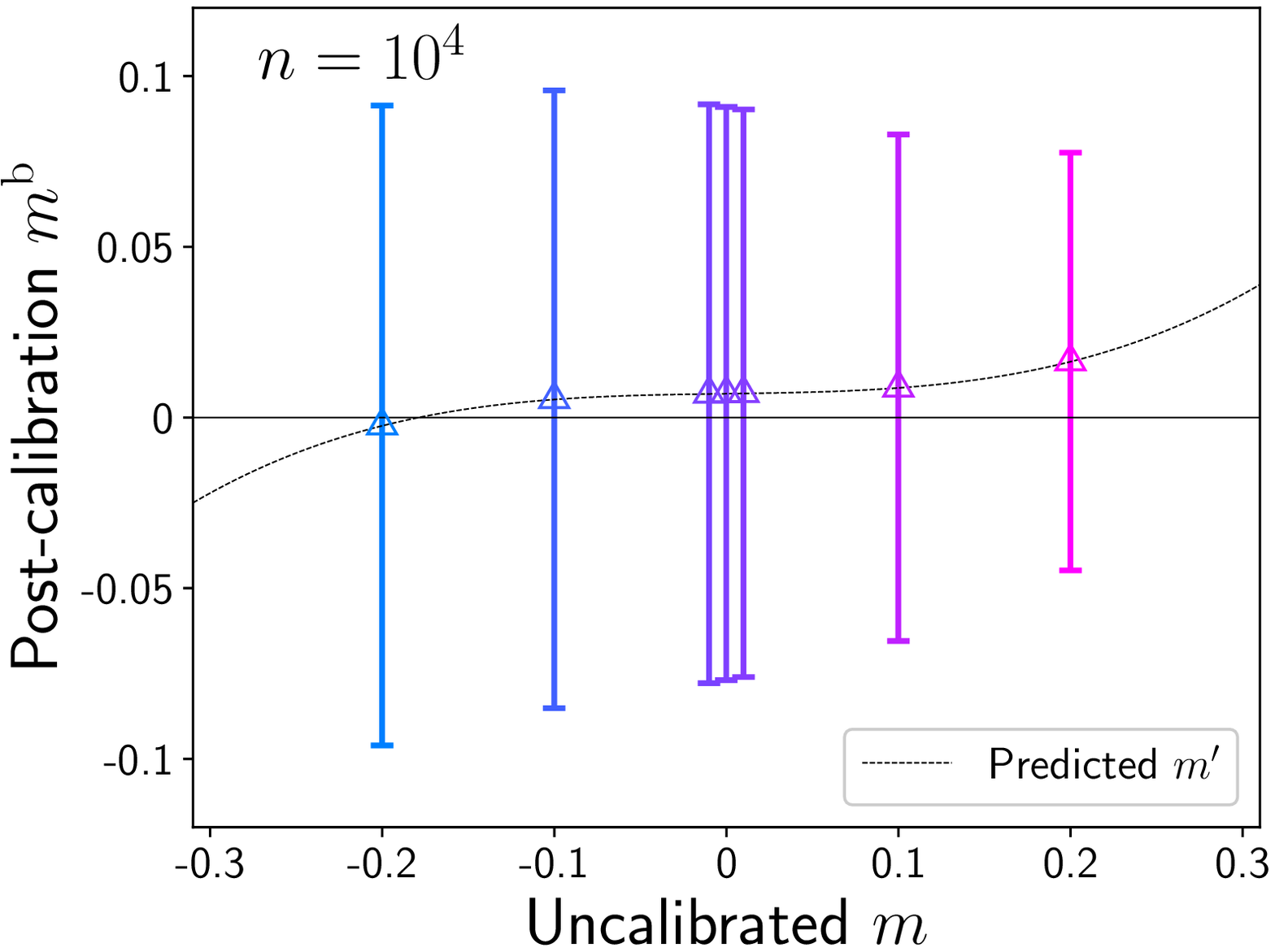}
	\includegraphics[scale=0.53]{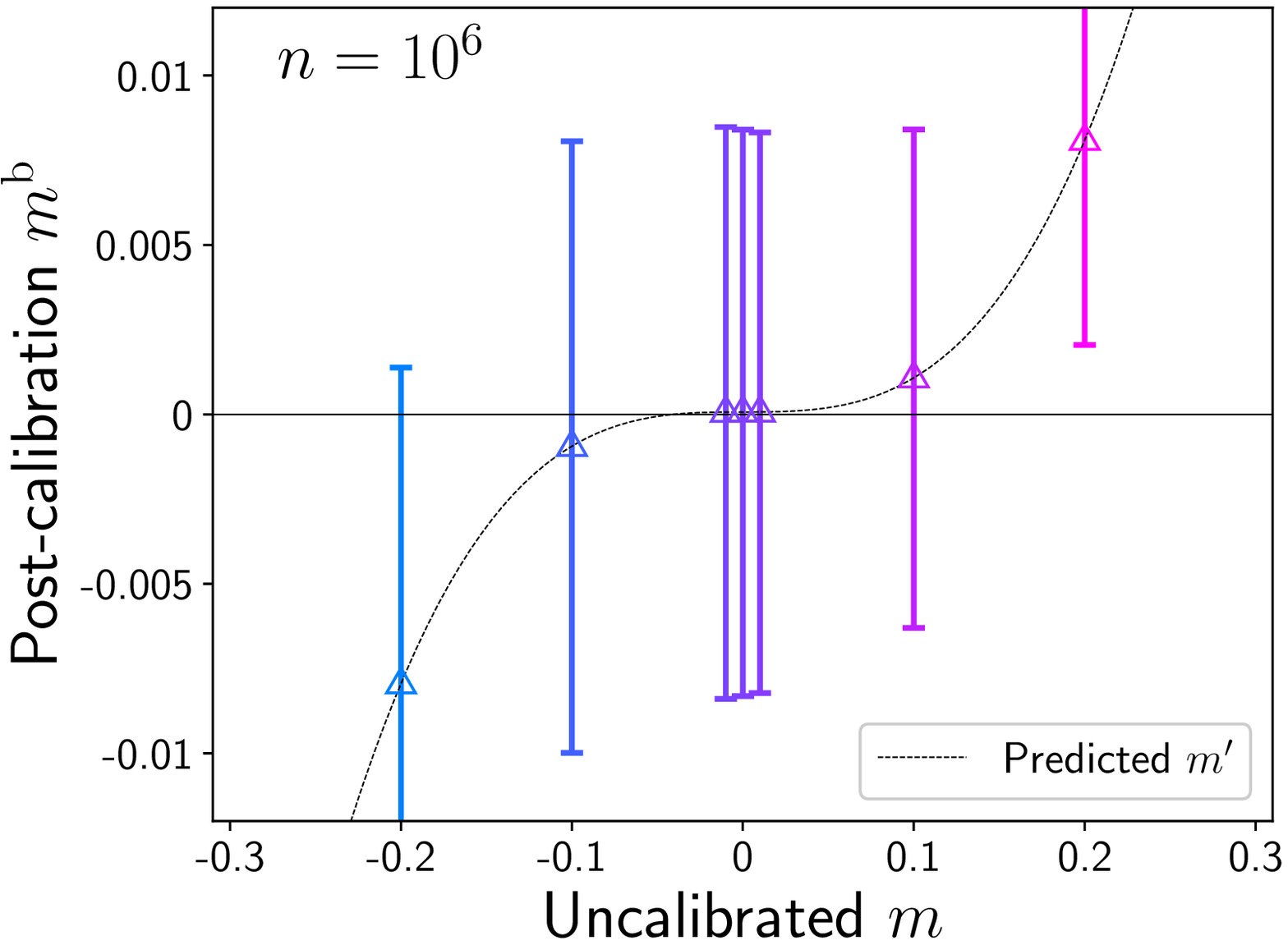}
	\caption{As the bottom-left panels of \Figref{bias_correction_illustration_1e4} and \Figref{bias_correction_illustration_1e6}, except showing the Bayesian-inspired correction instead of the first-order frequentist correction. The dashed line shows the prediction for the first-order frequentist correction, to illustrate that the results of the Bayesian-inspired correction are functionally identical to it.}
	\Figlabel{bias_correction_illustration_bayesian}
\end{figure*}

We show the post-correction distributions of $m$ and $c$ for this method in \Figref{bias_correction_illustration_bayesian}. The residual bias and scatter of this correction are essentially the same as the first-order correction we previously proposed based on a frequentist analysis, and so we see no reason to recommend this correction, with its more complicated form, over the simpler correction in \Eqref{bias_correction}.

\subsection{Conditional Calibration}
\Seclabel{cond_calibration}

One might consider calibrating a method for its multiplicative bias only if the bias is measured to be sufficiently large (for instance, if its absolute value is greater than the standard error in its measurement), and otherwise not calibrating. This approach can be consider in whole an alternative means of calibration, where the calibration equations are expressed as, e.g.:
\begin{align}
	\hat{\g}' = \begin{cases}
	        	\left(\hat{\g}-\hat{c}\right) & \left| \hat{m} \right| < m\sbr{t} \\
	        	\left(\hat{\g}-\hat{c}\right)\left(1-\hat{m}+\hat{m}^2\right) & \mathrm{otherwise,}
				\end{cases}
\end{align}
where $m\sbr{t}$ is the threshold for choosing whether or not to calibrate for the multiplicative bias.

Consider the case of a method with true bias $m$, measured bias $\hat{m}$, error in measured bias $\std{m}$, and threshold for not calibrating $m\sbr{t}=\std{m}$. If we choose to calibrate (using a first-order calibration) no matter the value of $\hat{m}$, then the resulting post-calibration $m'$ will follow the statistics presented in \Secref{calib_stats}, where the post-calibration bias is generally well-centred on zero. If we choose not to calibrate in the cases where $\left|\hat{m}\right|<m\sbr{t}$, then this is effectively using a more complicated calibration procedure with a choice in it, resulting in the post-calibration bias:
\begin{align}
	m^{c} = \begin{cases}
	        	m & \left| \hat{m} \right| < m\sbr{t} \\
	        	m' & \mathrm{otherwise.}
	        \end{cases}
\end{align}

\begin{figure}
	\includegraphics[scale=0.45]{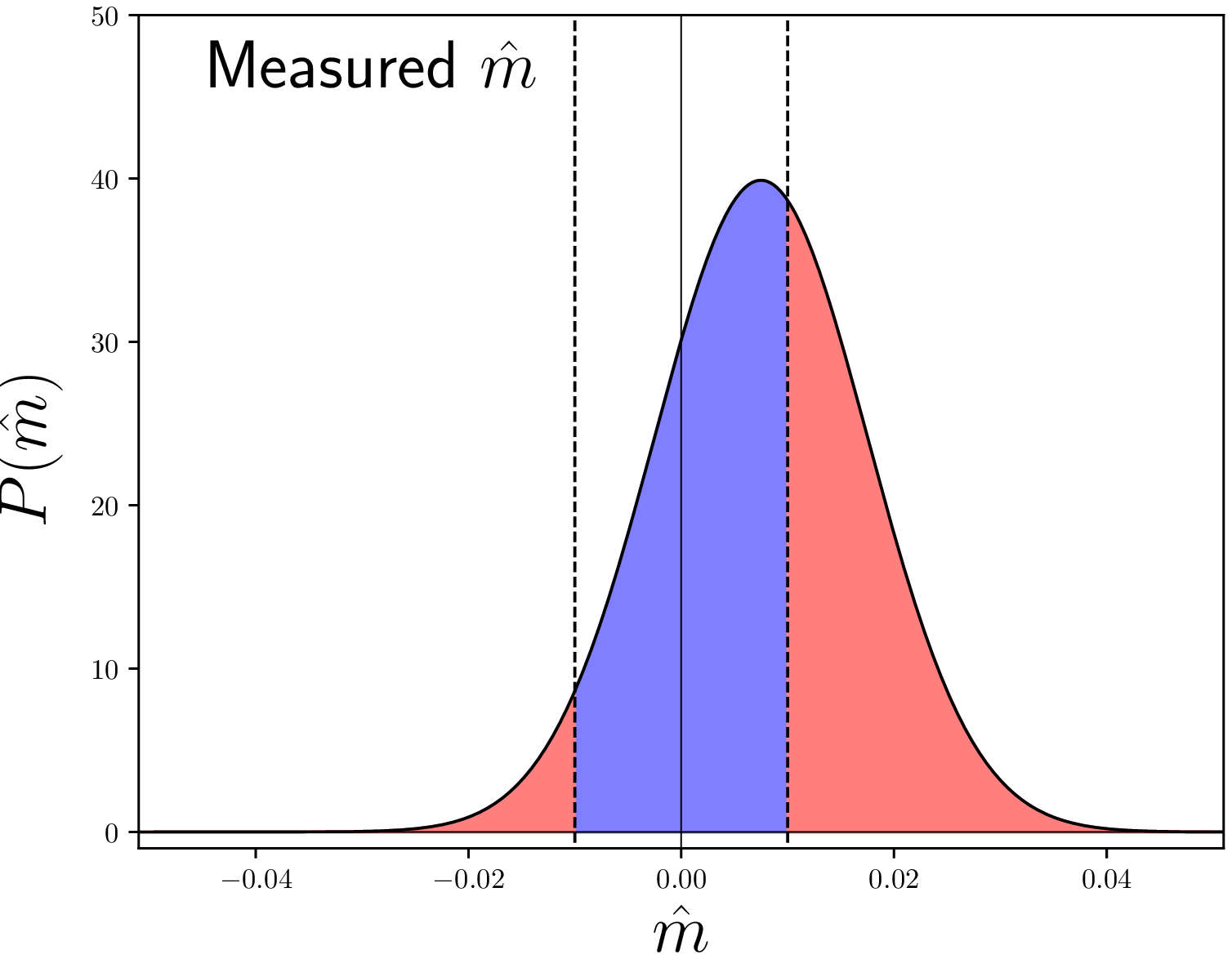}
	\includegraphics[scale=0.45]{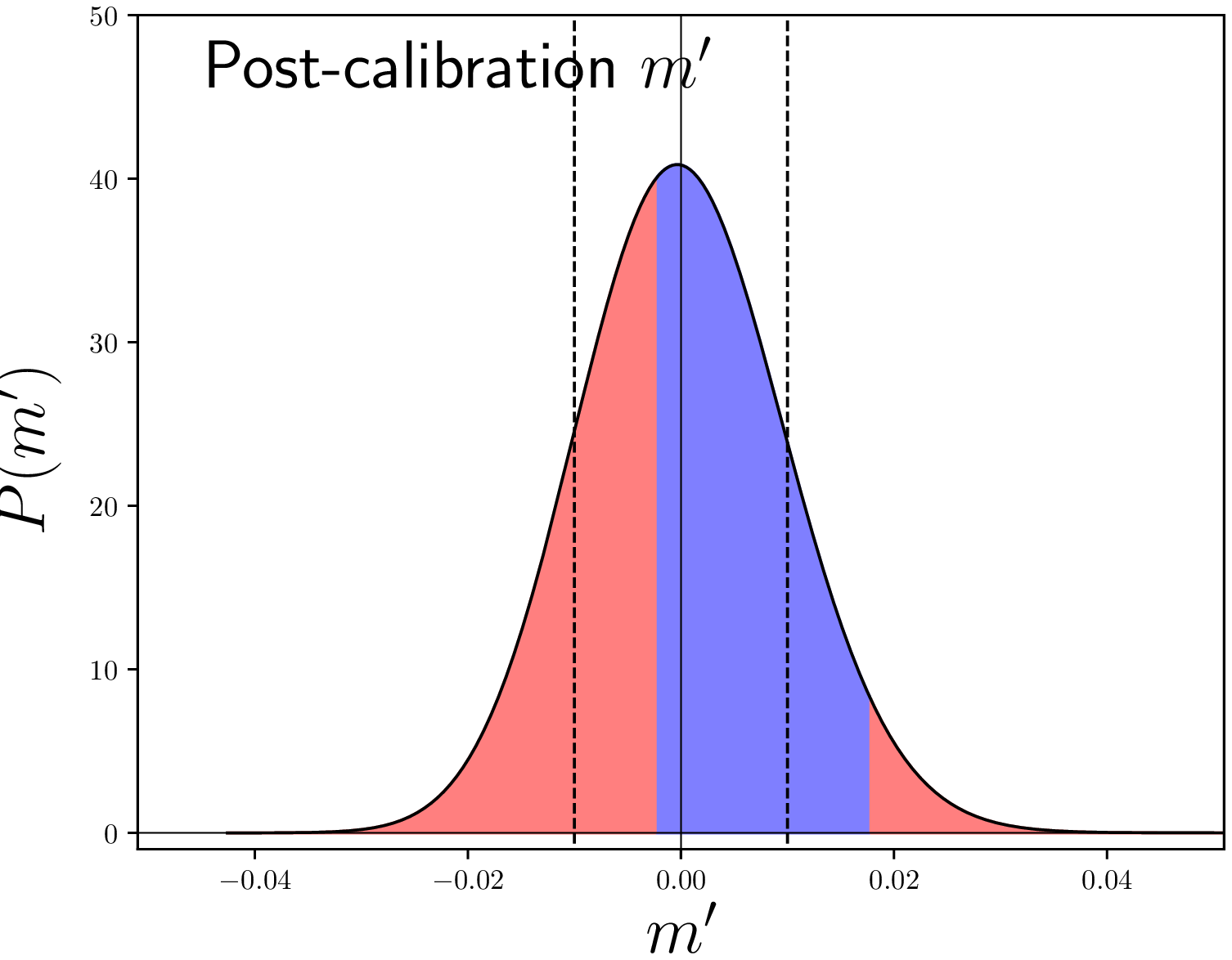}
	\includegraphics[scale=0.45]{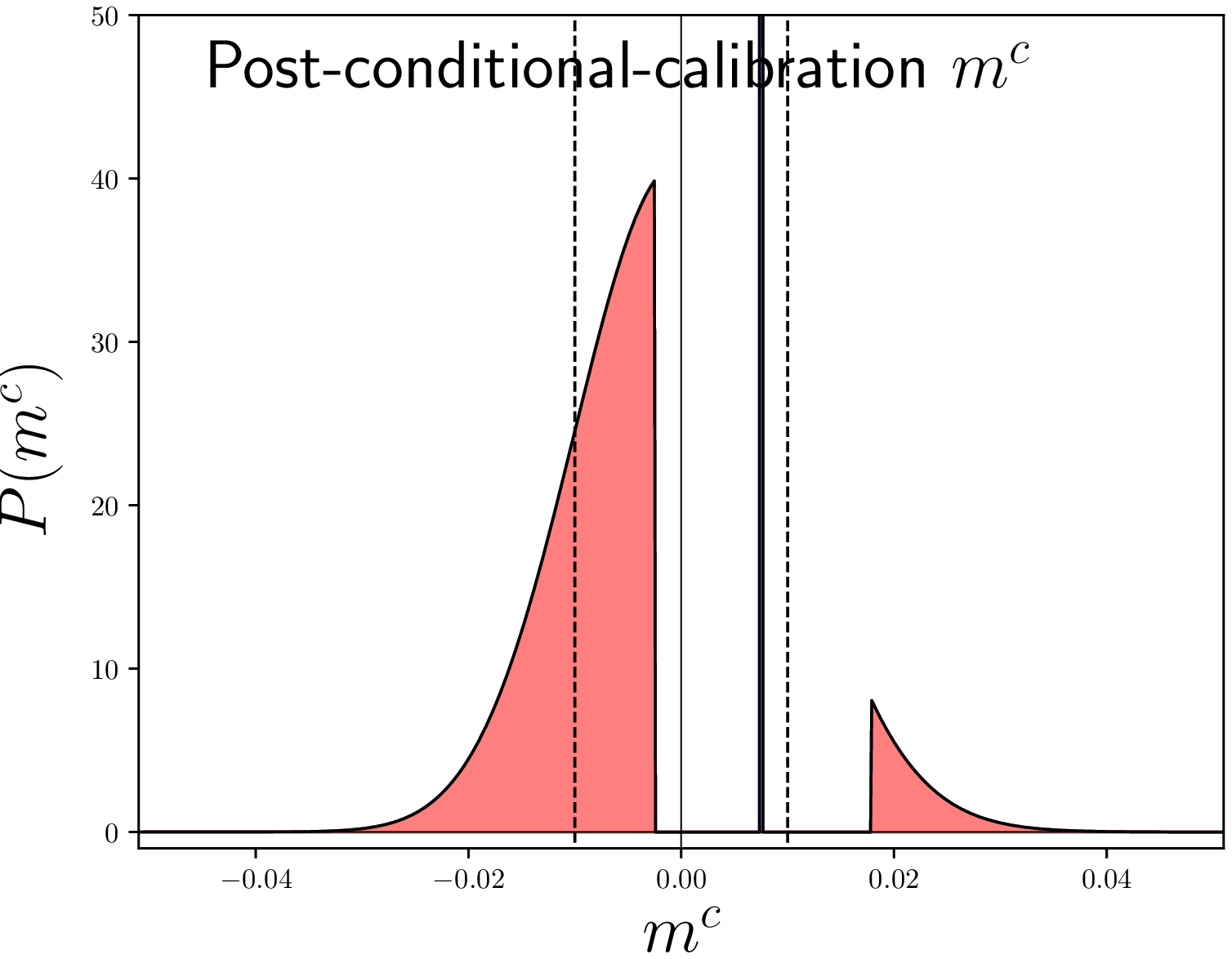}
	\caption{Distributions of the post-calibration $m$ for a case of conditional calibration, where a method is only calibrated if $\left|\hat{m}\right| < \std{m}$, using example values of true bias $m=0.0075$ and $\std{m}=0.01$. The top panel shows the distribution of the measured $\hat{m}$, the middle panel shows the distribution of the post-calibration $m'$ using the standard first-order calibration, and the bottom panel shows the post-calibration distribution of the conditionally-calibrated $m^c$. The shaded blue regions represent the values of $\hat{m}$ for which calibration would not be performed, and the red regions represent the values for which it would be performed. In the bottom panel, the blue region manifests as a delta function.}
	\Figlabel{calibration_choice_example}
\end{figure}

We present an example probability distribution of $m^{c}$ in \Figref{calibration_choice_example}, using example values of true bias $m=0.0075$ and $\std{m}=0.01$. The result of a procedure like this is that the region of the probability distribution surrounding $m$ is compressed to a delta function. If $m$ is close enough to zero ($\lesssim m\sbr{t}/2$), this results in the bias typically being closer to zero, but if $m$ is too far from zero, this makes the bias slightly further from zero on average (due to the asymmetry of the distribution in the region of $m$). If $m\sbr{t}$ is chosen proportional to $\std{m}$, this means that there will rarely be a case where choosing not to calibrate will result in an overall benefit, considering all possible values for $m$. This is especially the case if we consider the complications introduced by the more-complicated posterior distribution for $m^{c}$, which can complicate further analysis using these results.

It is also reasonable to consider that the threshold might be chosen based on external requirements, as is the case in the Euclid mission, which has a requirement of $\left|m\right| < 2\times10^{-3}$, and the value of $m$ will be measured to much greater precision than this. In this case, if $\left|m\right|$ is well below the threshold value, calibration will still provide a benefit over not calibrating except in cases where $\left|m\right| \ll \std{m}$ (in which case the slight increase of the scatter in post-calibration $m'$ will outweigh the benefit of bringing $\expec{m'}$ closer to zero), but this scenario cannot be known to be the case. Thus, always calibrating is the preferred option.

This logic can be extended to a decision never to calibrate; the difference being that the chosen threshold will be infinity rather than $\std{m}$. However, in this case, the posterior distribution of $m$ will simply be the result of its measurement, and with this simpler distribution, it is possible that further calculations which rely on the shear estimates might be able to incorporate this knowledge in a proper analysis. This may prove beneficial, but a full analysis of this will have to be investigated in the context of the measurements one wishes to use the shear estimates for and is thus beyond the scope of this paper. 

\section{Discussion and Conclusions}
\Seclabel{conclusions}

We have shown in this paper that while the additive component of shear bias, $c$, can be perfectly removed through calibration on average, this is not the case for the multiplicative bias, $m$. The non-linear nature of its effects on shear measurements result in error on its measurement not propagating linearly through calibration, resulting in a residual bias on average. It is possible to reduce this residual bias iteratively, but each iteration will increase the scatter of the post-calibration bias, slowly diverging.

The approach taken in this paper is purely frequentist, assuming no prior information on the bias. However, if priors can be reasonably approximated by Gaussian distributions, they can be incorporated into the approach here relatively easily by simply using the means and standard deviations of the posterior distributions of the bias parameters in the calculations shown here. Additionally, although we do make the assumption of Gaussianity for the probability distribution functions of the bias parameters and shape measurement error, most of the calculations presented here do not rely heavily on this assumption; for the first-order calibration, it is only an issue in the calculation of the post-correction scatter in the bias parameters and in the projections of their error based on the design of the calibration data.

We found that a Bayesian approach can also be taken to correcting for bias, and we present the resulting likelihood function for $P(\hat{\g}|\g,\hat{m},\hat{c})$ in \Eqref{likelihood_P_ghat}. While this likelihood function can be used to propose a different bias correction from the frequentist approach, we find no advantage to doing so when it comes to the residual bias or scatter.

A further issue is that calibration of this form assumes that the bias in the calibration dataset matches the bias in observations, and that there are no issues if subsets of the dataset remain biased after calibration (eg. if the bias scales with galaxy size and only the mean bias is corrected, galaxies in certain size ranges will remain biased after correction). This can be mitigated by binning galaxies by parameters that affect bias and then calibrating bins separately, but this will require a significantly larger amount of calibration data. The determination of the best method to handle this issue is beyond the scope of this paper, and it will require further research.

Through analysing the mean and scatter of the residual bias after a first- and second-order correction, we find that the first-order correction described in \Eqref{bias_correction} provides the optimal balance of reducing mean residual bias and not overly increasing its scatter if using a simulation size of $\sim 10^6$ galaxies or more per bin and if we expect $|m| \lesssim 0.1$, but the second-order calibration will be needed for larger $m$ values or smaller simulation sizes. We also find that while it might seem intuitively reasonable to not perform calibration if the measured bias is under a certain threshold, this rarely provides significant benefit and complicates the posterior probability distribution of the bias enough that it is unlikely to be worthwhile. We thus recommend that calibration always be performed, even if it results in a nearly-trivial correction, unless further analysis can gracefully use non-calibrated results with knowledge of the posterior distribution of the bias, in which case the choice should be made beforehand to never calibrate.

The predicted mean residual bias after this correction is shown in \Eqref{post_correction_known_biases} and the variance of the residual bias is shown in \Eqref{post_correction_vars}. As these predictions rely on knowledge of the actual bias, we present unbiased estimators of these which rely only on measured values in \Eqref{post_correction_known_biases_estimate} and \Eqref{post_correction_known_vars_estimate} for the first-order correction, and similar calculations can be used to determine similar expressions for the second-order correction. The required size of the calibration set for a given bias target can be estimated through \Eqref{n_sims_estimate}.

\section*{Acknowledgments}

The authors acknowledge useful conversations with Alex Hall and Lance Miller, and the feedback of the anonymous referees for this paper. BRG and ANT thank the UK Space Agency for funding, and ANT thanks the Royal Society for a Wolfson Research Merit Award.

\bibliographystyle{../Magnification_Method/mn2e}
\setlength{\bibhang}{2.0em}
\setlength\labelwidth{0.0em}
\bibliography{../Magnification_Method/full_bib}

\appendix

\section{Expanded Calculations}
\Applabel{exp_calcs}

To calculate the residual bias in our estimate of $g$, we start with \Eqref{g_bias_first_step} and take the expectation value:
\begin{align}
	\expec{\hat{\g}'} &= \expec{(1 - \delta_m  + m\delta_m + \delta_m^2 + m^3 + 2m^2\delta_m + m\delta_m^2) \g} \\ \nonumber
				 &\;\;\;\;{}+\expec{(\delta_{\g}-\delta_c)(1 - m - \delta_m + m^2 + 2m\delta_m + \delta_m^2)} \\ \nonumber
	             &= (1 - \expec{\delta_m}  + m\expec{\delta_m} + \expec{\delta_m^2} + m^3 + 2m^2\expec{\delta_m} + m\expec{\delta_m^2}) \g \\ \nonumber
				 &\;\;\;\;{}+\expec{\delta_{\g}}-\expec{\delta_c} - m\left(\expec{\delta_{\g}}-\expec{\delta_c}\right) -
					\expec{\delta_m}\left(\expec{\delta_{\g}}-\expec{\delta_c}\right) \\ \nonumber
				 &\;\;\;\;{}+ m^2\left(\expec{\delta_{\g}}-\expec{\delta_c}\right) + 2m\expec{\delta_m}\left(\expec{\delta_{\g}}-\expec{\delta_c}\right) \\ \nonumber
					&\;\;\;\;{}+ \expec{\delta_m^2}\left(\expec{\delta_{\g}}-\expec{\delta_c}\right)
\end{align}
Using the substitutions in \Eqref{expec_values}, this can be simplified to give \Eqref{gp_expec_value}.

The full forms for the expected variances on $m'$, $c'$, and $g'$ are
\begin{align}
	\Eqlabel{post_correction_vars_full}
	\var{m'} = \var{m}&\bigg[1 - 2m\bigg( 1 + \frac{3}{2}m - 2m^2 - 2\var{m} \\ \nonumber
		&\;\;- 2m^3 - m\var{m}\bigg) + 2\var{m}\bigg]\mathrm{,} \\ \nonumber
	\var{c'} = \var{c}&\bigg[1 - 2m\bigg(1-\frac{3}{2}m + m^2 + 3\var{m} \\ \nonumber
		&\;\;- \frac{1}{2}m^3 - 3m\var{m}\bigg) \\ \nonumber
		&\;\;\;\;\;\;\;\;\;\;\;\;+ 3\var{m}\bigg(1 + \var{m}\bigg)\bigg]\mathrm{,} \\ \nonumber
	\var{\g'} = \var{\g}&\bigg[1 - 2m\bigg(1-\frac{3}{2}m + m^2 + 3\var{m} \\ \nonumber
		&\;\;- \frac{1}{2}m^3 - 3m\var{m}\bigg) \\ \nonumber
		&\;\;\;\;\;\;\;\;\;\;\;\;+ 3\var{m}\bigg(1 + \var{m}\bigg)\bigg]\mathrm{.}
\end{align}

The full form for the expected known bias of $m''$ is
\begin{align}
	\Eqlabel{second_known_bias_full}
	\expec{m''} &= \var{m}\bigg[\var{m} + 3m^2 - 7m\var{m} - 5m^3 \\ \nonumber
	&\;\;\;\;\;\;\;\;\;\;\;\;\;\;\;- 8\sigma^4\left[m\right] + 8m^4 \bigg] - m^6 \mathrm{.}
\end{align}

The first-order approximation for \Eqref{likelihood_P_ghat_rewritten} is
\begin{align}
	\Eqlabel{likelihood_P_ghat_first_order}
	&P'(\hat{\g}|\g,\hat{m},\hat{c}) = A \exp\Bigg[\\ \nonumber
	&\;\;\;\; \frac{-1}{2(\var{c}+\var{g})}\Bigg(\left((1+\hat{m})\g + \hat{c} - \hat{\g} \right)^2\left(1-\frac{g^2\var{m}}{\var{c}+\var{g}}\right)  + \\ \nonumber
	&\;\;\;\;\;\;\;\;\left(-2\left(\var{c}+\var{g}\right) + g^2\var{m}\right) \Bigg) \Bigg] \\ \nonumber
	&\;\;\;\;= A \exp\Bigg[ -\frac{1}{2}\frac{ (1+\hat{m})^2 + \var{m} - (\hat{\g} - \hat{c})^2 \frac{\var{m}}{\var{c} + \var{g}} }{\var{c}+\var{g}} \\ \nonumber
	&\;\;\;\; \times \left( \g - \frac{(\hat{\g} - \hat{c})(1+\hat{m})}{(1+\hat{m})^2 + \var{m} - (\hat{\g}-\hat{c})^2 \var{m}/(\var{c} + \var{g}) }\right)^2 \Bigg] \mathrm{.}
\end{align}

\section{Supplementary Data}
\Applabel{supp_data}

\begin{table*}
	
	\caption{The values for the data plotted in \Figref{bias_correction_illustration_1e4}, showing the measured pre-calibration bias parameters and the actual post-calibration bias parameters for a dataset of size $n=10^4$.}

	\begin{center}
\begin{tabular}{rrrrrrrrrrrrrr}
$m$ & $c$ & $\overline{\hat{m}}$ & $\overline{m'}$ & $\overline{m''}$ & $\std{\hat{m}}$& $\std{m'}$ & $\std{m''}$ & $\overline{\hat{c}}$ & $\overline{c'}$ & $\overline{c''}$ & $\std{\hat{c}}$ & $\std{c'}$ & $\std{c''}$ \\ \hline
 -0.200 & -0.1 & -0.2001 & -0.0024 & 0.0011 & 0.0834 & 0.0937 & 0.1063 & -0.1000 & -0.0000 & -0.0000 & 0.0025 & 0.0031 & 0.0031 \\
 -0.200 & 0.0 & -0.2001 & -0.0024 & 0.0011 & 0.0834 & 0.0937 & 0.1063 & 0.0000 & -0.0000 & -0.0000 & 0.0025 & 0.0031 & 0.0031 \\
 -0.200 & 0.1 & -0.2001 & -0.0024 & 0.0011 & 0.0834 & 0.0937 & 0.1063 & 0.1000 & -0.0000 & -0.0000 & 0.0025 & 0.0031 & 0.0031 \\
 -0.100 & -0.1 & -0.1001 & 0.0053 & 0.0004 & 0.0834 & 0.0905 & 0.0940 & -0.1000 & -0.0000 & -0.0000 & 0.0025 & 0.0028 & 0.0028 \\
 -0.100 & 0.0 & -0.1001 & 0.0053 & 0.0004 & 0.0834 & 0.0905 & 0.0940 & 0.0000 & -0.0000 & -0.0000 & 0.0025 & 0.0028 & 0.0028 \\
 -0.100 & 0.1 & -0.1001 & 0.0053 & 0.0004 & 0.0834 & 0.0905 & 0.0940 & 0.1000 & -0.0000 & -0.0000 & 0.0025 & 0.0028 & 0.0028 \\
 -0.010 & -0.1 & -0.0101 & 0.0069 & 0.0001 & 0.0834 & 0.0847 & 0.0849 & -0.1000 & -0.0000 & -0.0000 & 0.0025 & 0.0026 & 0.0025 \\
 -0.010 & 0.0 & -0.0101 & 0.0069 & 0.0001 & 0.0834 & 0.0847 & 0.0849 & 0.0000 & -0.0000 & -0.0000 & 0.0025 & 0.0026 & 0.0025 \\
 -0.010 & 0.1 & -0.0101 & 0.0069 & 0.0001 & 0.0834 & 0.0847 & 0.0849 & 0.1000 & -0.0000 & -0.0000 & 0.0025 & 0.0026 & 0.0025 \\
 0.000 & -0.1 & -0.0001 & 0.0070 & 0.0001 & 0.0834 & 0.0839 & 0.0840 & -0.1000 & -0.0000 & -0.0000 & 0.0025 & 0.0025 & 0.0025 \\
 0.000 & 0.0 & -0.0001 & 0.0070 & 0.0001 & 0.0834 & 0.0839 & 0.0840 & 0.0000 & -0.0000 & -0.0000 & 0.0025 & 0.0025 & 0.0025 \\
 0.000 & 0.1 & -0.0001 & 0.0070 & 0.0001 & 0.0834 & 0.0839 & 0.0840 & 0.1000 & -0.0000 & -0.0000 & 0.0025 & 0.0025 & 0.0025 \\
 0.010 & -0.1 & 0.0099 & 0.0071 & 0.0001 & 0.0834 & 0.0831 & 0.0831 & -0.1000 & -0.0000 & -0.0000 & 0.0025 & 0.0025 & 0.0025 \\
 0.010 & 0.0 & 0.0099 & 0.0071 & 0.0001 & 0.0834 & 0.0831 & 0.0831 & 0.0000 & -0.0000 & -0.0000 & 0.0025 & 0.0025 & 0.0025 \\
 0.010 & 0.1 & 0.0099 & 0.0071 & 0.0001 & 0.0834 & 0.0831 & 0.0831 & 0.1000 & -0.0000 & -0.0000 & 0.0025 & 0.0025 & 0.0025 \\
 0.100 & -0.1 & 0.0999 & 0.0087 & 0.0002 & 0.0834 & 0.0742 & 0.0761 & -0.1000 & -0.0000 & -0.0000 & 0.0025 & 0.0023 & 0.0023 \\
 0.100 & 0.0 & 0.0999 & 0.0087 & 0.0002 & 0.0834 & 0.0742 & 0.0761 & 0.0000 & -0.0000 & -0.0000 & 0.0025 & 0.0023 & 0.0023 \\
 0.100 & 0.1 & 0.0999 & 0.0087 & 0.0002 & 0.0834 & 0.0742 & 0.0761 & 0.1000 & -0.0000 & -0.0000 & 0.0025 & 0.0023 & 0.0023 \\
 0.200 & -0.1 & 0.1999 & 0.0164 & 0.0004 & 0.0834 & 0.0612 & 0.0698 & -0.1000 & -0.0000 & -0.0000 & 0.0025 & 0.0021 & 0.0021 \\
 0.200 & 0.0 & 0.1999 & 0.0164 & 0.0004 & 0.0834 & 0.0612 & 0.0698 & 0.0000 & -0.0000 & -0.0000 & 0.0025 & 0.0021 & 0.0021 \\
 0.200 & 0.1 & 0.1999 & 0.0164 & 0.0004 & 0.0834 & 0.0612 & 0.0698 & 0.1000 & -0.0000 & -0.0000 & 0.0025 & 0.0021 & 0.0021 \\
\end{tabular}
\end{center}

	\Tablabel{table_data_1e4}

\end{table*}

\begin{table*}
	
	\caption{The values for the data plotted in \Figref{bias_correction_illustration_1e6}, showing the measured pre-calibration bias parameters and the actual post-calibration bias parameters for a dataset of size $n=10^6$.}

	\begin{center}
\begin{tabular}{rrrrrrrrrrrrrr}
$m$ & $c$ & $\overline{\hat{m}}$ & $\overline{m'}$ & $\overline{m''}$ & $\std{\hat{m}}$& $\std{m'}$ & $\std{m''}$ & $\overline{\hat{c}}$ & $\overline{c'}$ & $\overline{c''}$ & $\std{\hat{c}}$ & $\std{c'}$ & $\std{c''}$ \\ \hline
 -0.200 & -0.1 & -0.2000 & -0.0080 & -0.0001 & 0.0084 & 0.0094 & 0.0104 & -0.1000 & -0.0000 & -0.0000 & 0.0002 & 0.0003 & 0.0003 \\
 -0.200 & 0.0 & -0.2000 & -0.0080 & -0.0001 & 0.0084 & 0.0094 & 0.0104 & 0.0000 & -0.0000 & -0.0000 & 0.0002 & 0.0003 & 0.0003 \\
 -0.200 & 0.1 & -0.2000 & -0.0080 & -0.0001 & 0.0084 & 0.0094 & 0.0104 & 0.1000 & -0.0000 & -0.0000 & 0.0002 & 0.0003 & 0.0003 \\
 -0.100 & -0.1 & -0.1000 & -0.0010 & -0.0000 & 0.0084 & 0.0090 & 0.0093 & -0.1000 & -0.0000 & -0.0000 & 0.0002 & 0.0003 & 0.0003 \\
 -0.100 & 0.0 & -0.1000 & -0.0010 & -0.0000 & 0.0084 & 0.0090 & 0.0093 & 0.0000 & -0.0000 & -0.0000 & 0.0002 & 0.0003 & 0.0003 \\
 -0.100 & 0.1 & -0.1000 & -0.0010 & -0.0000 & 0.0084 & 0.0090 & 0.0093 & 0.1000 & -0.0000 & -0.0000 & 0.0002 & 0.0003 & 0.0003 \\
 -0.010 & -0.1 & -0.0100 & 0.0000 & -0.0000 & 0.0084 & 0.0084 & 0.0084 & -0.1000 & -0.0000 & -0.0000 & 0.0002 & 0.0003 & 0.0003 \\
 -0.010 & 0.0 & -0.0100 & 0.0000 & -0.0000 & 0.0084 & 0.0084 & 0.0084 & 0.0000 & -0.0000 & -0.0000 & 0.0002 & 0.0003 & 0.0003 \\
 -0.010 & 0.1 & -0.0100 & 0.0000 & -0.0000 & 0.0084 & 0.0084 & 0.0084 & 0.1000 & -0.0000 & -0.0000 & 0.0002 & 0.0003 & 0.0003 \\
 0.000 & -0.1 & 0.0000 & 0.0000 & -0.0000 & 0.0084 & 0.0084 & 0.0084 & -0.1000 & -0.0000 & -0.0000 & 0.0002 & 0.0002 & 0.0002 \\
 0.000 & 0.0 & 0.0000 & 0.0000 & -0.0000 & 0.0084 & 0.0084 & 0.0084 & 0.0000 & -0.0000 & -0.0000 & 0.0002 & 0.0002 & 0.0002 \\
 0.000 & 0.1 & 0.0000 & 0.0000 & -0.0000 & 0.0084 & 0.0084 & 0.0084 & 0.1000 & -0.0000 & -0.0000 & 0.0002 & 0.0002 & 0.0002 \\
 0.010 & -0.1 & 0.0100 & 0.0000 & -0.0000 & 0.0084 & 0.0083 & 0.0083 & -0.1000 & -0.0000 & -0.0000 & 0.0002 & 0.0002 & 0.0002 \\
 0.010 & 0.0 & 0.0100 & 0.0000 & -0.0000 & 0.0084 & 0.0083 & 0.0083 & 0.0000 & -0.0000 & -0.0000 & 0.0002 & 0.0002 & 0.0002 \\
 0.010 & 0.1 & 0.0100 & 0.0000 & -0.0000 & 0.0084 & 0.0083 & 0.0083 & 0.1000 & -0.0000 & -0.0000 & 0.0002 & 0.0002 & 0.0002 \\
 0.100 & -0.1 & 0.1000 & 0.0011 & -0.0000 & 0.0084 & 0.0074 & 0.0076 & -0.1000 & -0.0000 & -0.0000 & 0.0002 & 0.0002 & 0.0002 \\
 0.100 & 0.0 & 0.1000 & 0.0011 & -0.0000 & 0.0084 & 0.0074 & 0.0076 & 0.0000 & -0.0000 & -0.0000 & 0.0002 & 0.0002 & 0.0002 \\
 0.100 & 0.1 & 0.1000 & 0.0011 & -0.0000 & 0.0084 & 0.0074 & 0.0076 & 0.1000 & -0.0000 & -0.0000 & 0.0002 & 0.0002 & 0.0002 \\
 0.200 & -0.1 & 0.2000 & 0.0081 & -0.0001 & 0.0084 & 0.0060 & 0.0070 & -0.1000 & -0.0000 & -0.0000 & 0.0002 & 0.0002 & 0.0002 \\
 0.200 & 0.0 & 0.2000 & 0.0081 & -0.0001 & 0.0084 & 0.0060 & 0.0070 & 0.0000 & -0.0000 & -0.0000 & 0.0002 & 0.0002 & 0.0002 \\
 0.200 & 0.1 & 0.2000 & 0.0081 & -0.0001 & 0.0084 & 0.0060 & 0.0070 & 0.1000 & -0.0000 & -0.0000 & 0.0002 & 0.0002 & 0.0002 \\
\end{tabular}
\end{center}

	\Tablabel{table_data_1e6}

\end{table*}

\Tabref{table_data_1e4} shows the values for the data plotted in \Figref{bias_correction_illustration_1e4}, and \Tabref{table_data_1e6} shows the same for \Figref{bias_correction_illustration_1e6}.

\label{lastpage}

\end{document}